\documentclass[twocolumn]{aastex701}
\usepackage{multirow}
\usepackage{enumerate}
\usepackage{mhchem}
\shorttitle{Icy Moons \& Rings Often Outshine Their Planets}
\shortauthors{Hopper et al.}

\begin{document}

\title{Exomoons and Exorings with the Habitable Worlds Observatory. III. Icy Moons and Rings Often Outshine Their Planets}

\author[0009-0004-6318-6874]{Sam Hopper}
\affiliation{Department of Astronomy, University of Michigan, Ann Arbor, MI 48109, USA}
\email[show]{samuelhoppr@gmail.com}

\author[0000-0002-9521-9798]{Mary Anne Limbach}
\affiliation{Department of Astronomy, University of Michigan, Ann Arbor, MI 48109, USA}
\email{mlimbach@umich.edu}

\author[0000-0001-5831-9530]
{Rachel Bowens-Rubin}
\affiliation{Eureka Scientific Inc., 2542 Delmar Ave., Suite 100, Oakland, CA 94602, USA}
\affiliation{Department of Astronomy, University of Michigan, Ann Arbor, MI 48109, USA}
\email{rbowens-rubin@mit.edu}

\begin{abstract}

 If we could observe the Solar System’s giant planets with the upcoming \textit{Habitable Worlds Observatory} (HWO), the dominant sources of reflected near-infrared light would not always be the planets themselves. In fact, the brightest sources are split roughly evenly between the giants and their icy companions: at the locations of the giant planets, the dominant point sources would be Jupiter, Saturn’s rings, Uranus, and Triton (Neptune’s icy moon). Motivated by these Solar System cases in which rings or moons outshine their host planets, we investigate the detectability of analogous exomoons and exorings with HWO. We compute the expected signal-to-noise ratios ($S/N$) for moons and rings with the same albedos as their Solar System counterparts over a range of sizes, system distances, and star to planet+moon or star to planet+ring separations. We find that reflected light from Triton-analog exomoons and Saturn-analog exorings is detectable with HWO over a broad parameter space. In the nearest systems, icy Titan-sized moons are detectable at star-moon separations of $\sim$2\,AU and icy Earth-sized moons out to $\sim$6\,AU. Moreover, Saturn-like exorings are likely to be the brightest point sources in many extrasolar systems, capable of greatly outshining any planets, while remaining potentially detectable at system distances out to 100\,pc. These results highlight the need for further work to develop methods for identifying and disentangling exoring and exomoon signatures from exoplanet signals in HWO observations, where these sources will be blended and spatially unresolved.

\end{abstract}


\section{Introduction}\label{sec:intro}

The search for moons and rings beyond our Solar System offers a critical next step in understanding planetary system architectures and evolution. Their occurrence and observable properties encode formation pathways and dynamical histories that are not accessible from planets alone.

Significant effort has gone into developing techniques to detect exomoons of directly imaged exoplanets with existing observatories, including astrometry \citep{2025arXiv250915304W}, radial velocity \citep{2018AJ....156..184V,2023AJ....165..113R}, and transits \citep{2024AJ....168...54L,2025arXiv251024575W}. Additional approaches monitor stellar light during planetary transits to identify moon signatures, such as outgassing from moons around short-orbit exoplanets \citep{Oza2019Sodium, Gebek2020Alkaline} and transit or TTV signals of a satellite \citep{2006A&A...450..395S,2007A&A...470..727S} or utilize microlensing detection \citep{2010A&A...520A..68L}. Despite this substantial community effort, only a handful of exomoon or circumplanetary candidates have been proposed, and none are confirmed \citep[e.g.,][]{Ben2014TRANSIT, Bennett_2014, Teachey2017HEK, Miyazaki2018MOA, Limbach2021}.

Exorings provide a complementary probe of circumplanetary environments. While protoplanetary disks, where moons presumably form, have been detected in several systems including J1407b \citep{2015ApJ...800..126K}, PDS~70c \citep{2021ApJ...916L...2B}, K2-33b \citep{2022ApJ...940L..30O}, AS~209 \citep{2022ApJ...934L..20B}, and YSES-1 \citep{2025Natur.643..938H}, rings around mature exoplanets have proven difficult to detect \citep{2020AJ....159..131P,2021AJ....161..202L,2025ApJ...980...39L}. Theory and observation suggest several formation channels, including collisions between minor bodies \citep{2015NatGe...8..686H}, tidal disruption or stripping of a minor body \citep{2010Natur.468..943C,2022Sci...377.1285W}, incomplete dissipation of a primordial circumplanetary disk \citep{2005Icar..175..111B}, and material ejection from an active moon \citep{1979Sci...204..951S}. Detecting exorings would directly constrain these processes and place the prevalence of Saturn ring analogs in context with our Solar System.

\begin{figure}[ht!]
\includegraphics[width=\columnwidth]{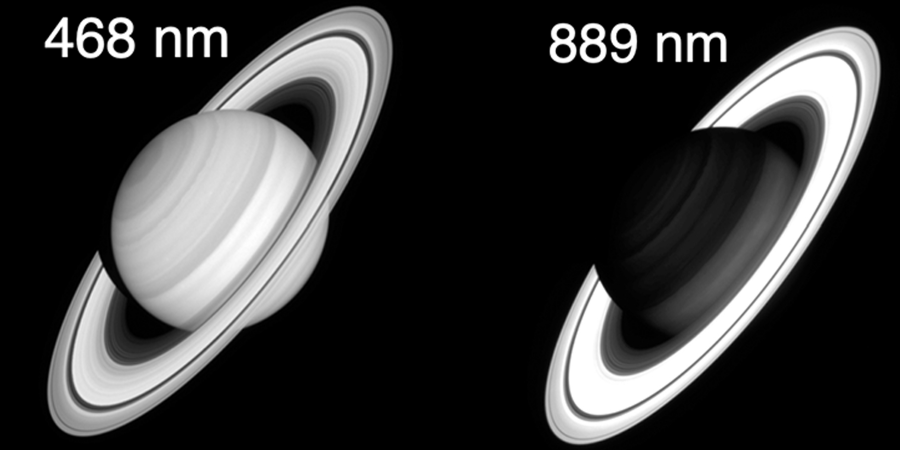}
    \caption{\textbf{Hubble photo of Saturn and its rings}. While both the planet and rings are visible at 468 nm \textit{(left)}, only the rings are visible at 889 nm \textit{(right)} due to the methane absorption in Saturn's atmosphere. This example illustrates how  Saturn's rings can outshine the planet by orders of magnitudes in portions of the near infrared. Image adapted from \cite{rs14061518}. 
\label{fig:SaturnRings}}
\end{figure}

\begin{figure}[ht!]
\includegraphics[width=\columnwidth]{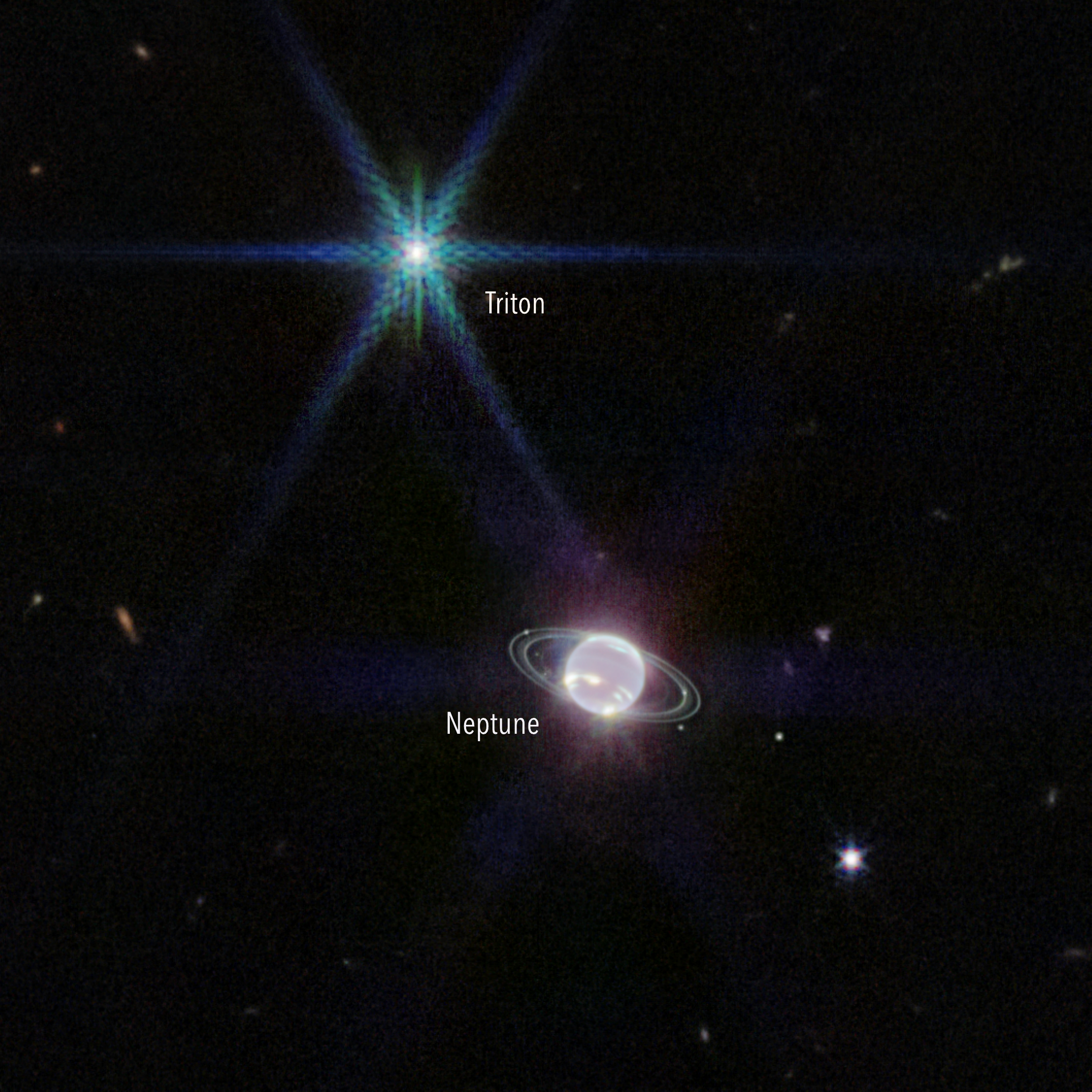}
    \caption{\textbf{JWST NIRCam image of Neptune and its moons}. Triton (bright blue point source in the upper left) outshines it's host planet Neptune in the near infrared (JWST NIRcam; Image Credit: NASA, ESA, CSA, and STScI).
\label{fig:ntNIRcam}}
\end{figure}

Within our Solar System, rings are ubiquitous. All four giant planets host rings \citep{1977Natur.267..328E,1986Natur.319..636H,1989Sci...246.1422S}, with Saturn’s being most prominent (see Figure~\ref{fig:SaturnRings}). Furthermore, minor bodies in the Solar System, such as Haumea, Chariklo, and Quoaur, also host rings \citep{Braga_Ribas_2014,2023Natur.614..239M, Ortiz_2017}. In the near-infrared, Saturn’s icy rings can outshine the planet by nearly $100\times$. This effect where Saturn's rings outshines the planet is observable at the wavelengths ($400 -2000$ \,nm) that will be accessible by forthcoming reflected-light facilities such as Roman Coronagraph Instrument, the Habitable Worlds Observatory, and Extremely Large Telescope/Planetary Camera and Spectrograph \citep{2021pdaa.book.....N,2021Msngr.182...38K,Follette2023, mennesson2020pavingwayfuturemissions}. Although this effect is most dramatic for rings because of their potentially large surface area, icy moons like Triton can also outshine their planets (see Figure~\ref{fig:ntNIRcam}). These considerations motivate a key question: \emph{as observatories capable of directly imaging reflected light from exoplanets come online, will exorings or exomoons dominate the observed signal and outshine their host planets?}

A major step forward in reflected light direct imaging will come from the Habitable Worlds Observatory \citep[HWO;][]{2021pdaa.book.....N}. HWO is envisioned as a space telescope operating from the ultraviolet through the near-infrared, with the primary objective of directly imaging and characterizing $\sim$25 potentially habitable worlds in the solar neighborhood. Its sensitivity to reflected light from terrestrial and giant planets raises the prospect of detecting, or being contaminated by, unresolved moon and ring contributions \citep{2004AsBio...4..400W,2004A&A...420.1153A,2018haex.bookE.158C}. Disentangling moon and ring signals from those of their host exoplanets will be critical for interpreting future datasets, since these companions can strongly impact measured fluxes and spectra. For example, unresolved satellites may bias atmospheric inferences, including producing biosignature false positives \citep{2014PNAS..111.6871R}. In most cases, it may not be apparent a priori that a detected point source is dominated by a moon or a ring rather than a planet. While deconfusion tools for reflected light direct imaging are emerging that can help resolve the orbits of different planets \citep{Hasler2025, Pogorelyuk2022}, little focus has been given to how rings and moons may be misidentified as planets or bias these methods, potentially increasing confusion rates and impacting inferred planet populations. It is therefore essential to understand the conditions under which rings/moons may dominate the measured signal. 

In this paper we explore the parameter space in which HWO could directly detect reflected light from moons and rings analogous to those in the Solar System. We examine two cases in which a moon and a ring in our Solar System outshine their host planets (Section \ref{sec:methods}). We then map the detectability of analogous companions across HWO’s anticipated operating regime (Section \ref{sec:results}) and discuss the implications for target prioritization, retrieval frameworks, and mitigation of false positives (Section \ref{sec:discuss}).

\section{Methods}\label{sec:methods}

We began our investigation by considering a hypothetical scenario in which HWO could image our Solar System from afar and possessed both the sensitivity and the field of view necessary to detect all of the planets, moons, and rings. Within this framework, we asked the following question: {\it Would the light detected by HWO from a given planet ever be significantly contaminated—or even dominated— by the light from that planet’s moons or rings?}

We find two stand-out cases within our Solar System: Neptune–Triton (Section 2.1) and Saturn–Rings (Section 2.2). The secondary component in both cases (Triton in the case of Neptune, and the ring system in the case of Saturn) outshines the host planet at some of the wavelengths where HWO is expected to operate. This is due to their ice-dominated compositions, which result in very high albedos and make them exceptionally reflective. Furthermore, the emitting area of Saturn’s ring system exceeds that of the planet itself across a wide range of viewing inclinations, further enhancing its contribution to the system’s total reflected light. Thus, if an HWO-like observatory could image our Solar System, in the near-IR, the resulting portrait would in fact comprise six planets, one moon (Triton), and Saturn's ring.\footnote{ While this is outside the focus of this paper, we note that the Earth’s Moon outshines Earth in the mid-infrared at $4.3\,\mu\mathrm{m}$ and $6.3\,\mu\mathrm{m}$ \citep{2011ApJ...741...51R}. HWO does not plan to operate at these wavelengths.}

\subsection{Triton and Neptune}
To explore the detectability of exomoons with HWO, we compiled spectral data from the literature on Neptune and Triton. This exercise follows the “Solar System as an exoplanetary system” framework explored in other cases in literature \citep[e.g.,][]{Martin2015SOLAR, Mayorga2021Transmission, madden_catalog_2018}, but with a specific focus on how moons and rings could influence what HWO will detect when imaging other planetary systems in reflected light.

The Neptune geometric albedo spectra ($\alpha_p$) used in this study is drawn from \cite{madden_catalog_2018} from observations obtained with the 1.5\,m Kanata Telescope at the Higashi-Hiroshima Observatory using the TRISPEC instrument (Triple Range Imager and Spectrometer with Polarimetry; \citealt{lundock_2009}). The original references provide details on the data reduction process and note specific spectral regions where the reflectance measurements may be less reliable due to contamination.

The Triton geometric albedo spectra ($\alpha_{m}$) used in this analysis is compiled from two sources. The reflectance data from 0.67 - 2.55\,$\mu\mathrm{m}$ were obtained with NASA's 3-meter Infrared Telescope Facility (IRTF) using the SpeX Infrared Spectrograph \citep{holler_surface_2016}. The data from 0.45 - 0.67\,$\mu\mathrm{m}$ were acquired with the 2.24\,$\mathrm{m}$ telescope at Mauna Kea Observatory using a Cassegrain Spectrograph and image-tube camera \citep{cruikshank_diameter_1979}. We note that the Triton spectra are relative reflectance spectra normalized to unity, and not absolutely calibrated geometric albedo spectra \citep{holler_surface_2016}. However, several independent photometric measurements have shown that Triton has a high geometric albedo at visible wavelengths \citep{1990GeoRL..17.1761N,1990Sci...250..419H} near unity where the measured Triton spectra are normalized to one. Because modest changes in the absolute normalization do not affect the conclusions of this work, we treat the normalized reflectance spectrum as an approximate geometric albedo spectrum for the calculations presented here.

At these wavelengths, the observed flux is dominated entirely by reflected stellar light, with negligible contribution from the respective object's thermal emissions. To assess the relative brightness of Triton and Neptune, we used the geometric albedo spectra described above to compute the flux ratio between the two bodies. This ratio quantifies where, spectrally, Triton is brighter than Neptune. We define the brightness of an object, $B_{m/p}$, as
\begin{equation}
B_{m/p} = I_*\alpha_{m/p} A
\end{equation}
where $I_*$ is the incident stellar flux, $\alpha_{m/p}$ is the object's geometric albedo, and $A$ is its projected area ($\pi r^2$, where $r$ is the radius of the object). Taking the ratio of the moon brightness to that of its host planet, we arrive at 
\begin{equation}\label{fluxratio}
\frac{F_m}{F_p} = \frac{B_{m}}{B_{p}} = \frac{\alpha_{m}r_{m}^2}{\alpha_{p}r_{p}^2}
\end{equation}
where $B_m$ and $B_p$ are the brightnesses of the moon and planet, respectively, and $r_m$ and $r_p$ are their corresponding radii. Since the moon will share the same orbital distance from the star as the planet, we assume $I_*$ is the same for both.

\begin{figure*}[]
\includegraphics[width=\textwidth]{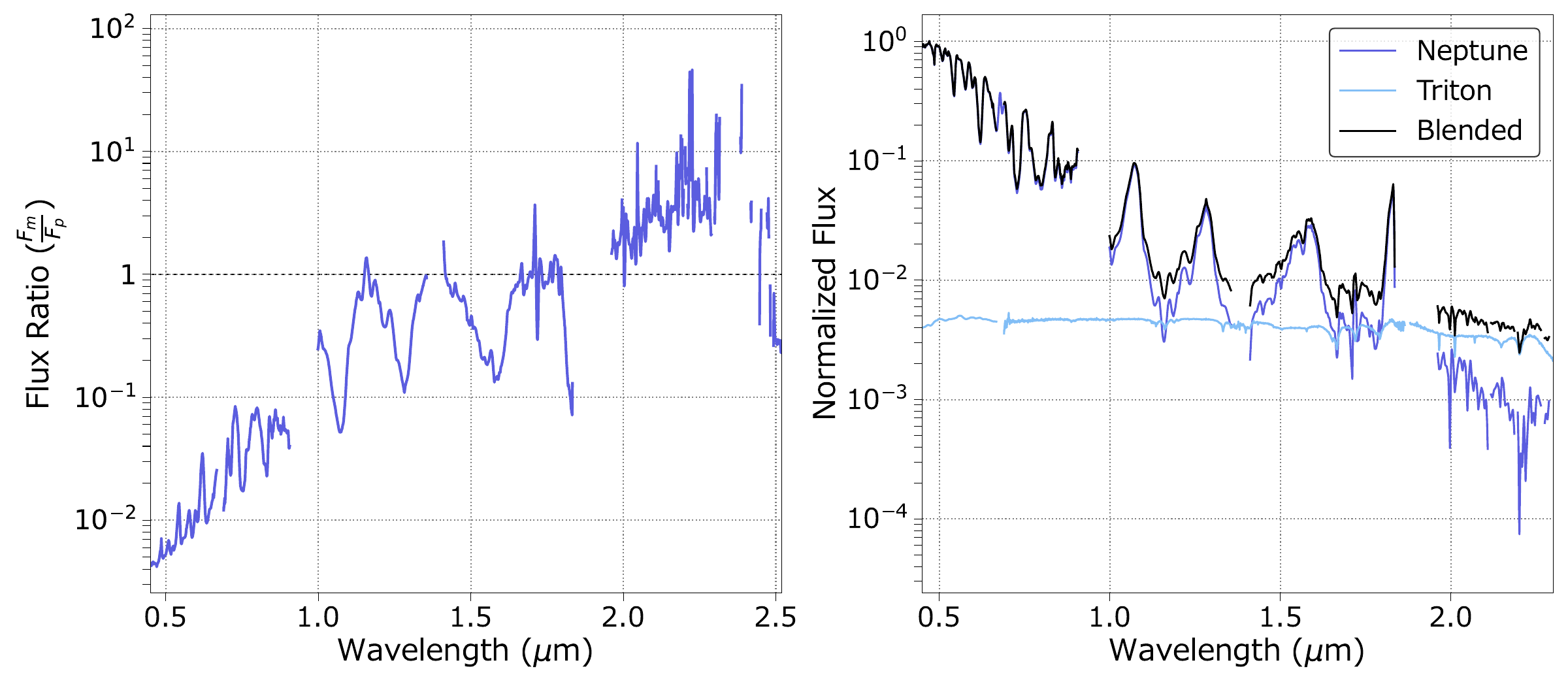}
    \caption{\textbf{Comparison of Triton and Neptune Brightnesses}. {\it Left panel:} The flux ratio
    of Triton compared to Neptune is shown in blue. The dashed line at Flux Ratio = 1 indicates equal brightness of Triton and Neptune. {\it Right panel:} The spectrum of Triton \textit{(light blue)}, spectrum of Neptune \textit{(blue)}, and their blended spectrum \textit{(black)} are shown normalized to a peak value of 1. Gaps in the spectra are due to unreliability in the data due to atmospheric contamination from Earth. Triton outshines Neptune by a factor of roughly 1.3$\times$ at 1.16 $\mu$m, 3.71$\times$ at $1.71\,\mu\mathrm{m}$ and up to 80$\times$ beyond $2\,\mu\mathrm{m}$ despite Neptune having an emitting area 18 times greater than Triton.}
\label{fig:tnfluxratio}
\end{figure*}

The left panel in Figure \ref{fig:tnfluxratio} highlights the flux ratio between Triton and Neptune. In this specific scenario, $r_m= 0.21\,R_{\oplus}$ \citep{Marques2022Constraints} and $r_p=3.86\,R_{\oplus}$ \citep{Seidelmann2007Report}. A flux ratio of 1 indicates equal brightness between the planet and its moon. Values greater than 1 signify that the moon is brighter than the planet at that wavelength.

Next we calculated the geometric albedo spectrum of a combined moon-planet system, which we refer to as a ``blended spectrum". This spectrum was obtained by summing the individual reflectance spectra of Triton and Neptune, such that
\begin{equation}\label{blend}
F = F_{m} + F_{p} = \alpha_{m}r_{m}^2 + \alpha_{p}r_{p}^2.
\end{equation} 
where $F$ is the total blended flux and $F_m$ and $F_p$ are the flux values specifically for the moon and planet, respectively. 

The right panel in Figure \ref{fig:tnfluxratio} presents the blended spectrum of Neptune and Triton, along with individual spectra of each object for comparison. The reflected light from Triton dominates the combined spectrum at $1.16\,\mu\mathrm{m}$, $1.71\,\mu\mathrm{m}$, and $>2\,\mu\mathrm{m}$. Triton outshines Neptune by a factor of 1.3$\times$ at $1.16\,\mu\mathrm{m}$, 3.71$\times$  at $1.71\,\mu\mathrm{m}$ and up to 80$\times$ beyond  $2\,\mu\mathrm{m}$. Despite the fact that the emitting area of Neptune is 18 times more than Triton, the regions where the moon is brighter correspond with regions of methane absorption in Neptune's atmosphere \citep{Fletcher2010Neptune}. Triton's composition is primarily composed of ices such as \ce{N2}, \ce{CH4}, \ce{CO}, and \ce{CO2} which contribute to its high reflectivity \citep{Cruikshank1993Ices}.

\subsection{Saturn and Rings}
For Saturn, we used geometric albedo data ($\alpha_p$) near its North Pole to avoid contamination from the ring system. 

These data originate from \cite{madden_catalog_2018}, who derived the geometric albedo spectra from observations obtained with the $1.5\,\mathrm{m}$ Kanata Telescope at Higashi-Hiroshima Observatory using the TRISPEC instrument \citep{lundock_2009}. We note that different regions of Saturn exhibit variations in albedo, cloud structure, and composition. While Saturn’s polar regions in particular can show temporal color variability due to changing cloud properties \citep{Sromovsky2021}, these effects are modest compared to the contribution of the ring signal considered here. Moreover, such variability is comparable to the diversity expected in exoplanet atmospheres more broadly. As a result, our use of a representative Saturn spectrum does not significantly impact the conclusions of this work.

For Saturn's rings, we used Bond albedo data ($\alpha_r$) of the B-ring obtained by the Cassini Spacecraft with the Visible and Infrared Mapping Spectrometer (VIMS) instrument and analyzed in \cite{ciarniello_cassini-vims_2019}. 
The data in \cite{ciarniello_cassini-vims_2019} used to derive the Bond albedo span a wide range of phase angles and viewing geometries. While reflectance does depend on these parameters, previous work has shown that these effects are most pronounced in specific regimes (e.g., opposition or extreme viewing geometries), and more modest across intermediate angles. We note that the Bond albedo is defined as the fraction of incident stellar radiation reflected by an object, integrated over all viewing angles (outgoing directions). As such, it does not correspond to a single phase angle, inclination or viewing geometry. We therefore adopt the Bond albedo as a simplified isotropic-scattering approximation and assume a ring viewing inclination of $i = 45^\circ$.

Following the same procedure used for Neptune and Triton, we define the brightness of a ring, $B_r$, as
\begin{equation}
B_r = \frac{1}{4}I_*\alpha_r A
\end{equation}
where $I_*$ is the incident stellar flux, $\alpha_r$ is the ring's Bond albedo, and $A$ is its projected area ($\pi (r_{o}^2 - r_{i}^2)\cos(i)$, where $r_o$ is the outer ring radius, $r_i$ is the inner ring radius, and $i$ is the ring viewing inclination angle). We take the ratio of the ring brightness to that of its host planet as
\begin{equation}\label{fluxratioring}
\frac{F_r}{F_p} = \frac{B_{r}}{B_{p}} = \frac{\alpha_{r}(r_{o}^2 - r_{i}^2)\cos(i)}{4\alpha_{p}r_{p}^2}.
\end{equation}

For this case, we used the size of the B-ring \citep[$r_o = 117507$\,km and $r_i =91975$\,km;][]{Murray2000Solar} for the ring radii, and $r_p = 0.84\,R_J$ \citep{Seidelmann2007Report}. When calculating the reflecting area of the rings, we assumed a ring viewing inclination of $i = 45^\circ$ relative to face-on or edge-on orientations, providing an intermediate value between the two extreme viewing inclinations.

\begin{figure*}[]
\includegraphics[width=\textwidth]{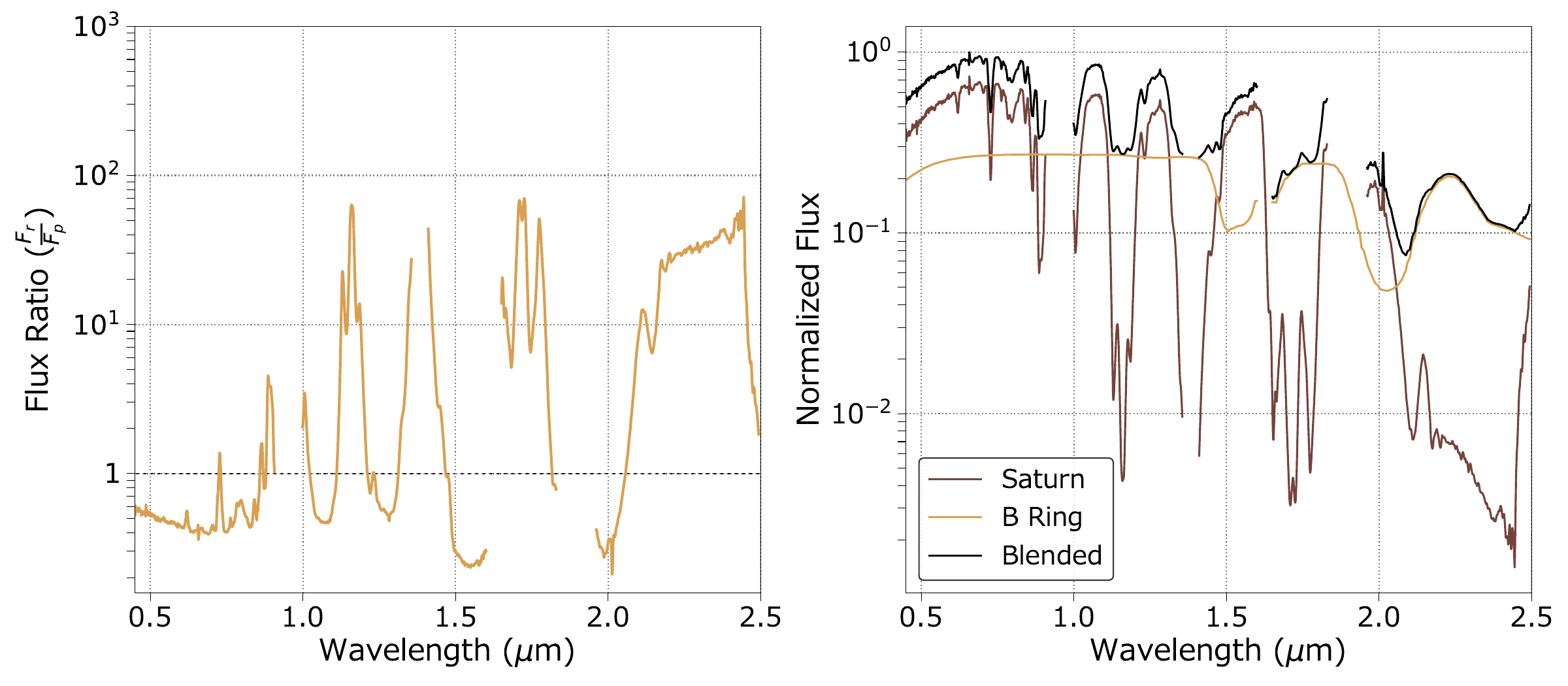}
    \caption{\textbf{Comparison of Saturn and its Rings Brightnesses}. {\it  Left panel:} The flux ratio of Saturn's B-ring relative to Saturn. The dashed line at Flux Ratio = 1 indicates equal brightness of B-ring and Saturn. {\it Right panel:} The spectrum of Saturn's B-ring \textit{(golden)}, spectrum of Saturn \textit{(brown)}, and their blended spectrum \textit{(black)} are shown normalized to a peak value of 1. At a fraction of wavelengths, the B-ring outshines Saturn, reaching more than $50\times$ brighter in certain regions of the spectrum.
\label{fig:srfluxratio}} 
\end{figure*}

 Figure \ref{fig:srfluxratio} (left panel) shows the flux ratio between Saturn and the B-ring. Similar to the Triton-Neptune case, a flux ratio of 1 indicates equal ring and planet brightness. A flux ratio greater than one indicates the ring is brighter than the planet. The blended Saturn-Rings spectrum is defined as
\begin{equation}\label{blendring}
F = F_{r} + F_{p} = \frac{1}{4}\alpha_{r}(r_{o}^2-r_i^2)\cos(i) + \alpha_{p}r_{p}^2.
\end{equation} 
where $F_r$ is the ring flux. 
 
 Figure \ref{fig:srfluxratio} (right panel) shows the blended spectrum of Saturn and the B-ring along with their individual spectra. The rings have a prominent influence on the blended spectrum at a substantial fraction of wavelengths and dominate in the near-IR (1.6 - 2.5\,$\mu\mathrm{m}$). We find that the wavelength where the rings outshine Saturn with the greatest ratio occurs around $2.44\,\mu\mathrm{m}$, where Saturn's B-ring reach up to 71 times brighter. Other regions where Saturn's rings are $>50\times$ brighter than the planet include $1.15 - 1.16\,\mu\mathrm{m}$, $1.70 - 1.73\,\mu\mathrm{m}$, $1.77\,\mu\mathrm{m}$, and $2.42 -2.44\,\mu\mathrm{m}$. A methane absorption feature exists in Saturn's spectrum at all of these wavelengths \citep{Coulter2022Jupiter, BAINES2005ATMOSPHERES}.

We note that, although Saturn's rings appear comparable in brightness to the planet when viewed from Earth, this is because they are typically seen nearly edge-on. In contrast, observations of exoplanetary systems will sample ring systems over a wide range of ring  viewing geometries, rather than being restricted to near edge-on configurations.

\subsection{HWO SNR on Triton and Saturn-ring Analogs}
Next we establish a method for computing HWO's expected signal-to-noise ratio (SNR) for various exorings and exomoons across a range of sizes and orbital separations. Our goal is to identify the parameter space in which HWO could detect moons and rings analogous to Triton and Saturn’s rings.

To estimate the SNR for a representative exoring or exomoon observation, we created a modified version of \texttt{MinimalCoron.py} in the \texttt{hwotools} python package to enable the use of a range of user-input distances (e.g., $2-30$\,pc).\footnote{\url{https://github.com/spacetelescope/hwo-tools}} \texttt{hwotools} is a python package that allows the user to run science simulations for HWO preparatory science and the \texttt{MinimalCoron.py} function allows one to simulate planet signal-to-noise ratios for HWO. We set the integration time to 25 hours and the telescope aperture to 8\,m. We used distances ranging from 2 - 30\,pc for exploring the exomoons \citep[corresponding to the range of distances to the HWO target stars;][]{2025PASP..137j4402T} and 2 - 100\,pc for the exorings (as Saturn-like exorings should be detectable at further distances than exoplanets).

Using the equation for contrast presented in \cite{Cahoy2010-zd, Cahoy2017-sq}, we computed the moon-to-star contrast to input for our SNR calculation as,
\begin{equation}\label{contrast_m}
   C(\phi) = \alpha_m \frac{ r_m^2}{ a^2}[\frac{\sin(\phi)+(\pi-\phi)\cos(\phi)}{\pi}]
\end{equation}
where $C$ is contrast, $a$ is orbital separation in AU, and $\phi$ is the moon (or planet) phase angle. In this work, we set the phase angle to $\phi=0$ (e.g., full phase), which then simplifies the contrast to
\begin{equation}\label{contrast_m}
   C = \alpha_m \frac{ r_m^2}{ a^2}
\end{equation}
Similarly, using the Bond albedo based contrast prescription presented in \cite{2004A&A...420.1153A}, we define the ring-to-star contrast as
\begin{equation}\label{contrast_r}
   C = \alpha_r \frac{\pi (r_o^2-r_i^2)\cdot\cos(i)}{4 \pi a^2} = \alpha_r\frac{(r_o^2-r_i^2)\cdot \frac{\sqrt{2}}{2}}{4a^2}.
\end{equation}

We adopted moon geometric albedos ($\alpha_m$) and ring Bond ($\alpha_r$) albedos corresponding to those of Triton and Saturn’s rings at a ring inclination of $i=45^\circ$. We chose $i = 45^\circ$ as a representative intermediate ring viewing inclination, avoiding extreme cases such as face-on or edge-on orientations. For Triton, we selected spectral regions within HWO’s expected wavelength coverage that maximize the contrast where Triton most strongly outshines Neptune. The value of the Triton-moon analog geometric albedo was $\alpha_m =0.83$ which corresponds to Triton's geometric albedo at $\lambda= 1.16$\,$\mu$m. For the Saturn-Ring system, although the rings outshine Saturn at nearly all wavelengths, we adopted  $\alpha_r = 0.66$ at a wavelength of $\lambda= 0.89$\,$\mu$m, where a methane absorption band in Saturn’s atmosphere \citep{rs14061518, Coulter2022Jupiter} further boosts the ring–planet contrast. While Saturn’s atmosphere shows stronger absorption features at longer wavelengths, we focused on shorter wavelengths where the stellar flux is higher and HWO will likely be more sensitive. Since Saturn’s rings remain bright relative to the planet across the spectrum, there is no compelling need to push to longer wavelengths. 
 
The \texttt{MinimalCoron.py} was set up to run in V-band, however since both of these albedo values we chose do not fall into the V-band, we adjusted the parameters in \texttt{MinimalCoron.py} to compute the the stellar flux assuming J = 3.65\,mag for the Triton-analog moons and I = 4.10\,mag for the Saturn Ring analogs \citep{Willmer2018Absolute}.\footnote{\url{https://irsa.ipac.caltech.edu/data/SPITZER/docs/dataanalysistools/tools/pet/magtojy/}} This assumes the host is a sun-like star.

We computed the contrast ratios over orbital separations from 0.7 - 5\,AU in the majority of cases. In the exorings case where the rings were equal to and larger than Saturn's rings, we computed the contrast ratios out to orbital separations of 50\,AU. All remaining simulation parameters match those in \texttt{MinimalCoron.py}. We do not include detector noise in our analysis, as it is expected to be negligible compared to other noise sources in this regime. A summary of the parameters used for the SNR calculations can be found in Table \ref{parameters}.

We caution the reader that the moon and ring contrasts are not calculated using identical scattering prescriptions. Specifically, the ring equation includes a factor of $1/4$ whereas the moon and planet contrast equations do not. This difference is because the references used here provide geometric albedos for the planets and moons, including Triton, but a Bond albedo for Saturn’s rings.

\begin{deluxetable}{lcc}
\tablecaption{Simulation Parameters Used in SNR Calculations}
\label{parameters}
\tablehead{
\colhead{Parameter} &
\colhead{Ring Case} &
\colhead{Moon Case}
}
\startdata
Distance (pc)                     & 2--100   & 2--30 \\
Orbital separation (AU)           & 0.7-- 50  & 0.7--15 \\
Bond albedo
                                  & 0.66     & -- \\
Geometric albedo
                                  & --       & 0.83 \\
Wavelength ($\mu$m)               & 0.89     & 1.16 \\
Stellar magnitude                 & $I=4.10$ & $J=3.65$ \\
\enddata
\end{deluxetable}

\section{Results}\label{sec:results}
\begin{figure*}[b]
\centering
\includegraphics[width=0.99\textwidth]{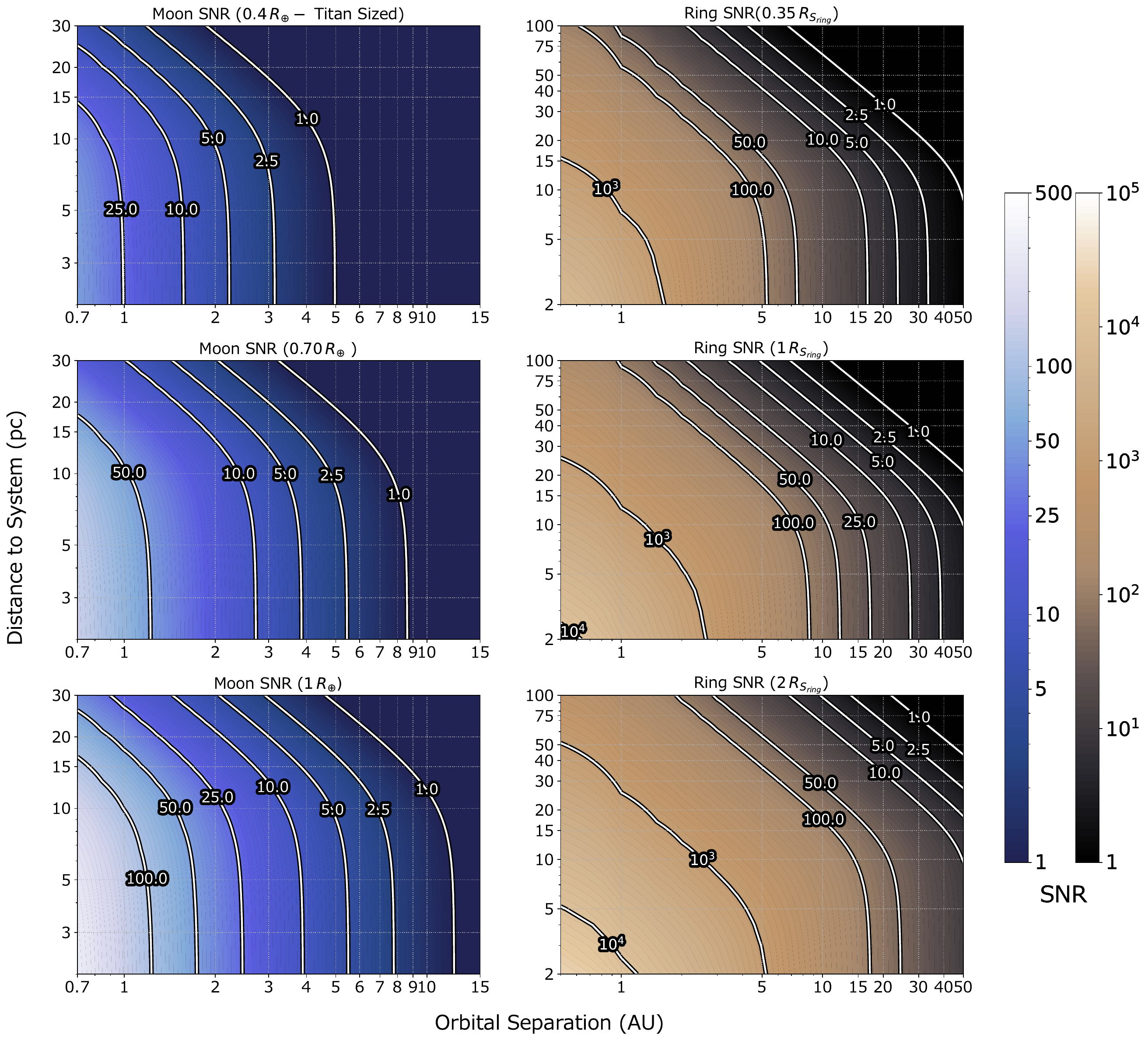}
    \caption{\textbf{Estimated SNR with HWO}. The estimated SNR values for exomoon and exorings were calculated across a variety of ring/moon sizes, orbital separation (e.g., the distance from the star to the planet+moon or planet+ring), 
    and system distances. The left column of 3 plots shows the SNR of exomoons of varying sizes from smallest (Titan radius) on the top to largest (Earth radius) on the bottom. The right column of 3 plots shows the SNR of exorings of varying sizes from smallest (B-ring radius) on the top to largest on the bottom (2$\times$ the radii of Saturn's A through D ring, or $2\times 1 R_{S_{ring}}$). The middle panel corresponds to Saturn-sized rings. Rings comparable to or larger than Saturn's are detectable to 5 $S/N$ across an impressively broad region of parameter space: they are detectable out to $a=5$\,AU to distances of 100\,pc, and at $a = 25$\,AU out to 20\,pc. Additionally, Titan-sized moons with Triton-like geometric albedos are detectable at 5 $S/N$ out to orbital separations of $\sim2$\,AU. Earth-sized moons with Triton-like geometric albedos are detectable out to $\sim5.5$\,AU for the closest systems.}
\label{fig:snr}
\end{figure*}

The results of our SNR calculations are shown in Figure \ref{fig:snr} for Triton-analogs (left-column) and Saturn's ring analogs (right-column) at various orbital separations and system distances. We define an object as marginally detectable if it has an SNR greater than 2.5, and as likely detectable if it has an SNR exceeding 5 (SNRs are marked on white lines in plot). Moons and rings are both detectable across portions of the parameter space we explore. Detectability increases with system proximity and with smaller star to planet+ring or planet+moon separations.

\subsection{Exomoon Detection Limits}
The left panels shows the SNR we may expect to achieve if we were to image moons that have geometric albedos like Triton. We selected 3 different sizes of moons to understand what size moon are we able to detect. The smallest size we chose was the size of Saturn's moon Titan \citep[$r= 0.4 R_{\oplus}$;][]{Murray2000Solar}. The ``mid-size" moon we selected had a radius of $r= 0.7 R_{\oplus}$. The ``large" moon selected was the radius of Earth. We note that, although Triton is brighter than Neptune, HWO will be not sensitive enough to detect either Neptune or Triton in reflected light. In general, exomoons will need to be somewhat larger and/or closer to their host star than Triton to be detectable with HWO, though not larger than some of the moons in our own Solar System.

For relatively nearby systems ($<$10\,pc), we find that a Titan-size moon with a Triton-like geometric albedo will likely be marginally detectable at star-moon separations inwards of 3\,AU and likely detectable inwards of 2\,AU. A $0.7\,R_{\oplus}$-size moon with a Triton-like geometric albedo will be marginally detectable out to 5\,AU and be likely detectable out to 4\,AU. An Earth-size moon with a Triton-like geometric albedo would remain marginally detectable out to 8\,AU and likely detectable out to 5.5\,AU. Inward of some distances, the occurrence rates of icy moons are likely to be low. However, examples in our Solar System, such as Snowball Earth, indicate that highly reflective terrestrial bodies at $\sim$1 AU are plausible; see discussion in Section~\ref{snowball-earth} for more detail.

\subsection{Exoring Detection Limits}
The right panels in Figure \ref{fig:snr} shows the SNR expected for imaging Saturn-like ring systems. Unlike moons, rings are much more readily detectable because their emitting area can be much larger than the planet, and their albedos, specifically for icy rings, are often higher than the planet's. While moons may also have significantly higher albedos, they cannot exceed the planet’s emitting area; rings benefit from both a high albedo and a large emitting area, making them potentially hundreds of times brighter than their host planet.

As with the exomoon detectability analysis, we considered three representative ring sizes. For the small case, we adopted the size of Saturn's B-ring. The mid-size ring used is a size equivalent to Saturn's A through D rings ($r_o = 136780$\,km and $r_i =  66900$\,km), which we define as 1 Saturn Ring ($1 R_{S_{ring}}$). For the large case, we used a ring system with a size equivalent to twice of Saturn's rings.  

Ring systems comparable in size to Saturn’s are detectable around stars as distant as 100\,pc inside separations of $<5$\,AU. For stars at distances within $20$\,pc, a ring system the same size as Saturn's could be detected within $<25-40$\,AU, although the outer working angle of HWO will be unlikely to enable constraints at such large separations for nearby systems.

Without any dedicated observations HWO will likely be able to provide occurrence rates for Saturn-like ring systems in the HWO star sample \citep{2024ApJS..272...30H,2025PASP..137j4402T}. This will be limited to the orbital separations enabled by the outer working angle of the coronagraph. Smaller ring systems comparable in size to just Saturn’s B ring are expected to be detectable across the HWO target star sample, which lies within $30$\,pc. A dedicated exoring survey of stars at larger distances (30-100\,pc) would enable occurrence rate constraints on exorings around planets at larger orbital separations. If icy exorings prove to be prevalent in nature, icy exorings could be among the most detectable point sources with HWO in extrasolar systems.

\section{Discussion}\label{sec:discuss}

\begin{figure*}
\includegraphics[width=\textwidth]{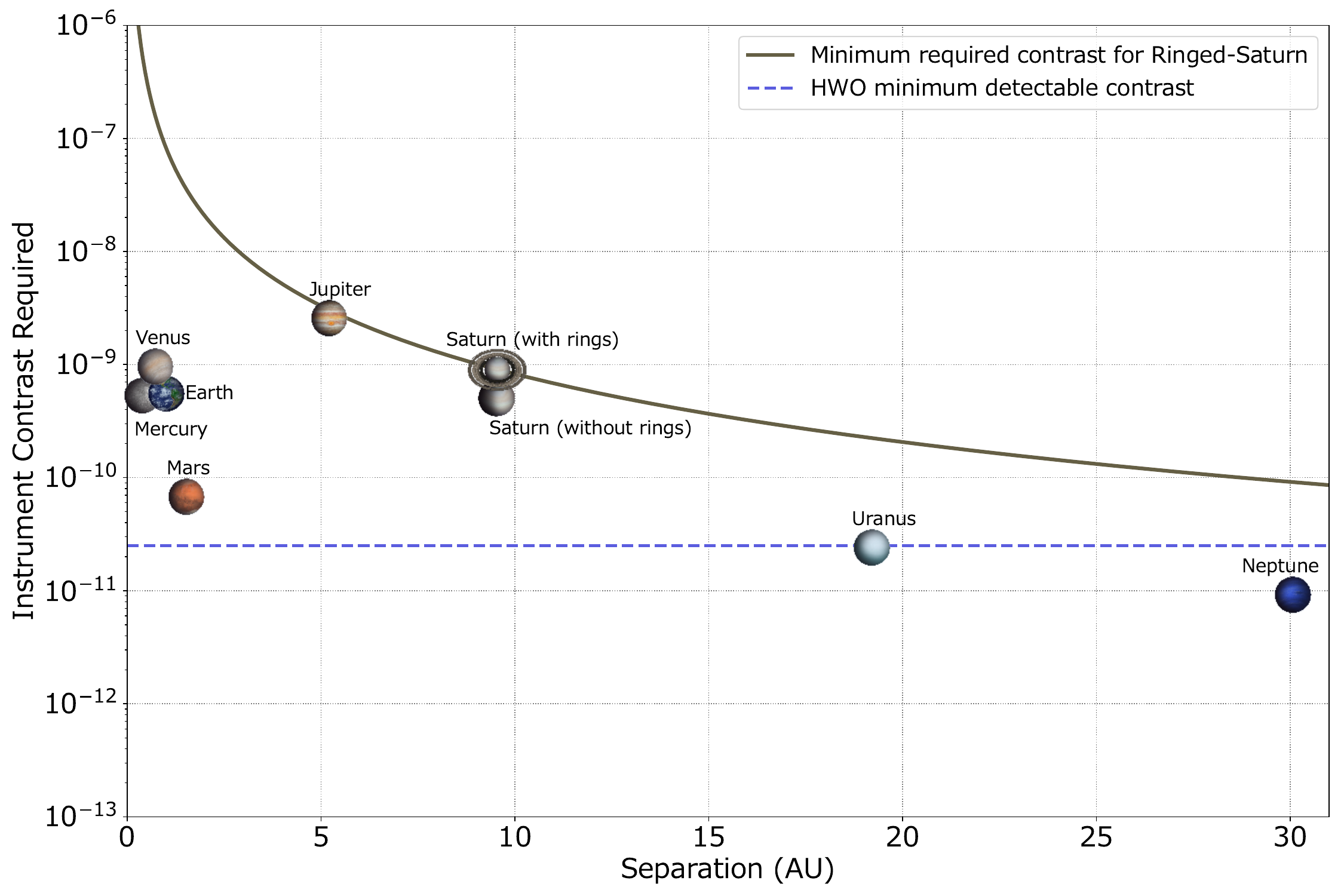}
    \caption{\textbf{The instrument contrast required for detecting the Solar System planets}. The planets, rings (with ring inclination set to $i=45^\circ$), and moons are denoted by their representative drawings. The minimum required contrast to detect a ringed-Saturn at various orbital separations from the sun is plotted as a brown solid line. Saturn is plotted twice to show its contrast with and without its rings. The minimum contrast for HWO (dotted blue line) is predicted to be around $2.5 \times 10^{-11}$ \citep{2024ApJS..272...30H}. Sources above the line would be detectable by HWO if between the inner and outer working angle and inside the field of view. 
\label{fig:contrastcurve}} 
\end{figure*}

Our findings highlight that even in our own Solar System the brightest point sources in reflected light are not always the planet themselves. Rather, we find that moons and rings can outshine their host planet and will be detectable with HWO. Saturn's rings, in particular, are the brightest point source in our Solar System (other than the Sun) at some wavelengths, even outshining Jupiter. Figure \ref{fig:contrastcurve} shows the contrast required to detect various Solar System planets as well as Saturn's rings placed at various orbital separation. 

We computed the contrast for the Solar System planets with Equation \ref{contrast_m} using a geometric albedo set to $\alpha_p = 0.3$ for each object as this a typical value for both terrestrial and gas giant planets in the visible \citep{Pearl_neptune_bond_albedo, HANEL1983262, PEARL199012, 2000GL012580}. The dashed blue horizontal line on the plot shows a plausible minimum detectable contrast for HWO of $2.5 \times 10^{-11}$ \citep{2024ApJS..272...30H}, although the actual limit will depend on the finalized observatory and instrument design. The plot shows the contrast ratio of the Sun to Saturn with and without its rings, as well the contrast of the sun to ringed-Saturn varying in orbital separations (brown line). We also continue to adopt a ring inclination of $i=45^\circ$. In the visible, Jupiter is slightly brighter than Saturn with its rings. However at some wavelengths in the near-IR, ringed-Saturn outshines Jupiter despite its larger orbital separation from the Sun, making it the brightest point source other than the star itself.

We note that Saturn’s rings are predominantly icy, whereas other known ring systems of giant planets \citep{2003A&A...412..305P, 1985Icar...63..312C, 1977Natur.267..328E,1986Natur.319..636H,1989Sci...246.1422S} and of some Solar System terrestrial bodies \citep{2014A&A...568A..79D, 2017Natur.550..219O} are rocky. In general, rocky rings and moons have much lower albedos than icy ones, implying that rocky objects would be more difficult to detect than the icy objects explored in this work. Given that a typical rocky geometric albedo is around $\sim0.05$, we expect rocky rings to be 1-2 order of magnitudes fainter than icy objects \citep{2004A&A...420.1153A}

\subsection{How Close Can Icy Moons Exist?}
\label{snowball-earth} 

We typically think about how far from a star liquid water can exist \citep[i.e., the outer extent of the habitable zone;][]{2013ApJ...765..131K} rather than how close icy bodies can persist. However, in the context of searching for icy exomoons, the opposite constraint becomes relevant. Icy bodies can in fact exist at relatively small orbital separations under the right conditions. As a familiar reference, we can consider periods in Earth’s history when it entered a ``Snowball Earth'' state during which the planetary albedo was significantly elevated \citep{2005JGRD..110.1111P}, potentially approaching values comparable to highly reflective icy bodies such as Triton. Geological and geochemical evidence indicates that Earth underwent at least two major Cryogenian glaciations, which may have approached global ice coverage and are widely described as ``Snowball Earth'' events \citep{1998Sci...281.1342H,2002TeNov..14..129H,2005GSAB..117.1181H,2010Sci...327.1241M}. Thus, over geological timescales, our own solar system provides an example of icy, terrestrial conditions persisting at $\sim$1\,AU.

We can estimate the minimum orbital separation at which it is possible for an icy moon to exist around a sun-like star. Assuming that stellar irradiation is the sole mechanism for heating, (e.g., negligible tidal and planetary heating), we estimate the orbital separation at which icy bodies could theoretically exist, using the equation
\begin{equation}
    a =\frac{R_{\star}}{2}\left(\frac{T_{eff}(1-\alpha)^{1/4}}{T_{eq}}\right)^2
\end{equation}
where $T_{eff}$ is the effective temperature, $R_{\star}$ is the stellar radius, and $T_{eq}$ is the equilibrium temperature of the moon. In this case, we used solar values ($T_{eff}=5778$\,K, $R_{\star}=1\,R_{\odot}$), an equilibrium temperature equal to the freezing point of water ($T_{eq}=273$\,K), and an albedo of $\alpha = 0.9$ (consistent with an icy body, such as Triton). Inputting these parameters, gives a minimum orbital separation for icy bodies to exist of only $a\sim0.33$\,AU. 

This estimate assumes radiative equilibrium only. In reality, surface temperatures depend on atmospheric properties, greenhouse effects, rotation, and heat transport \citep{2010ppc..book.....P}. On airless or low-pressure bodies, water ice is unstable to sublimation even well below 273\,K \citep{1961JGR....66.3033W,1999Icar..141..179V}, such that long-term stability of exposed ice generally requires temperatures $\lesssim 110$--150\,K \citep{2015Icar..255...58H}. For a high-albedo ($A \sim 0.9$) body, this corresponds to orbital separations of order $\sim$1-2\,AU around a Sun-like star. However, in the presence of an atmosphere, climate feedbacks can drive global glaciation even at $\sim$1\,AU, as evidenced by Snowball Earth episodes \citep{1998Sci...281.1342H,2002TeNov..14..129H}. We therefore consider icy moons to be plausible at small orbital separations and report detectability limits extending inward of 1\,AU.

\subsection{How can we distinguish a planet from a moon/ring?}
Although the reflected light from exomoons and exorings may be detectable with HWO, we generally do not expect to spatially resolve them from their host planet. Resolution may be possible in some edge cases, but this is expected to be rare \citep{Lazzoni_2022}. This highlights an important consideration: {\it When detecting an unresolved point source with HWO, how can we differentiate whether the point source is a moon, ring, or planet?} Given that there are cases in our own Solar System where moons and rings outshine their host planets, it is almost certain that similar situations will arise in exoplanetary systems observed by HWO.

For a given point source, differentiating the light between rings and a planet is an easier task than that for a moon and planet. Rings are likely to be a larger fraction of the reflected light, if not the dominant source of reflected light in a planetary system due to the generally larger surface areas. Measuring the polarization \citep{Lietzow_2023,Veenstra2025general} and phase curves \citep{2004A&A...420.1153A, Dyudina2005Phase} may prove critical in providing the constraints needed to disentangle the blended signal between the planets and their rings.

Future work is needed to develop methods to disentangle moon light blended with that of the planet's and build on the few techniques discussed in previous literature \citep{Your2024Paper, Agol2015CENTER}. One key path forward is the development of spectral retrieval codes capable of fitting two bodies simultaneously (e.g., planet+moon or planet+ring), providing a framework to identify cases in which a two-body fit is statistically favored, thereby indicating the presence of a moon or ring. In this work, we adopt full-phase geometric albedo spectra for the planets and moons, and use the Bond albedo for the rings at a single viewing inclination. These assumptions provide a reasonable first-order estimate of the effects of moons and rings but do not include the full dependence on phase angle, illumination geometry, and scattering. A more complete treatment of blended spectra will require exploring a broader range of geometries for the planet--moon--ring system.

Although this manuscript has highlighted the need to be able to differentiate if the light from a point source is dominated by a planet, a moon or a ring, implementing methods that will be able to conduct this differentiation is beyond the scope of this paper.

\section{Conclusion}
In this manuscript, we explored whether icy moons and rings could outshine their host planets leveraging Solar System analogs. Our findings show that both moons and rings can potentially outshine their hosts in reflected light at wavelengths accessible to HWO. Our main conclusions are summarized as follows:
\begin{itemize}
    \item HWO has the capacity to detect the reflected light from exomoons as small as Titan at star-moon separations of $\sim$2\,AU and Earth-sized moons out to $\sim$6\,AU for the closest star systems ($<10$\,pc) to 5 $S/N$. 
    \item Saturn-like and larger exorings remain detectable over a remarkably wide range of system architectures for HWO.  For systems at 100\,pc, Saturn-analog ring systems are detectable at star-ring separations of $a<5\,$AU. For systems at 20\,pc, they can be detected out to separations of $a<25$\,AU, provided they are not clipped by the field of view or by the outer working angle of the coronagraph.
    \item We cannot assume every point source we detect in exoplanetary systems with HWO is reflected light from an exoplanet. Exomoons and exorings can be significant sources of reflected light and can outshine their host planet. In our own Solar System, Saturn's rings are the brightest point source in the system at certain wavelengths, and Triton and the Moon outshine Neptune and Earth, respectively, across portions of the infrared. 
\end{itemize}

Our findings highlight the importance of developing techniques to separate the reflected-light signal between planets and their moons and rings. Because we find examples of moons and rings outshining their host planets within our own Solar System, we should expect that extrasolar systems on the HWO target list are likely to display the same features. To ensure HWO achieves its full scientific potential, it is imperative to develop tools that enable blended source spectral retrievals and properly account for the presence of rings and moons.

\section*{Acknowledgment}
We thank the anonymous referee for their constructive comments, which significantly improved the clarity, quality and content of this manuscript. S.H. thanks the University of Michigan's Honors Summer Fellowship Program for providing funding for this work.

\facilities{{\tt Habitable Worlds Observatory (HWO)}}.

\software{{\tt numpy.py} \citep{5725236}, {\tt matplotlib} \citep{Hunter2007Matplotlib}, {\tt Astropy} \citep{Robitaille2013Astropy}, {\tt pandas} \citep{pandas2025pandas}, {\tt SciPy} \citep{Virtanen2020SciPy} and {\tt hwo-tools} \citep{hwo-tools}.}

\bibliographystyle{aasjournalv7}
\bibliography{main}

@ARTICLE{1990GeoRL..17.1761N,
       author = {{Nelson}, R.~M. and {Burattl}, B.~J. and {Wallis}, B.~D. and {Smythe}, W.~D. and {Horn}, L.~J. and {Lane}, A.~L. and {Mayo}, M.~J. and {Simmons}, K.~E.},
        title = "{SPECTRAL GEOMETRIC ALBEDO AND BOLOMETRIC BOND ALBEDO OF NEPTUNE'S SATELLITE TRITON FROM VOYAGER OBSERVATIONS}",
      journal = {\grl},
         year = 1990,
        month = sep,
       volume = {17},
       number = {10},
        pages = {1761-1764},
          doi = {10.1029/GL017i010p01761},
       adsurl = {https://ui.adsabs.harvard.edu/abs/1990GeoRL..17.1761N}
}

@ARTICLE{1990Sci...250..419H,
       author = {{Hillier}, J. and {Helfenstein}, P. and {Verbiscer}, A. and {Veverka}, J. and {Brown}, R.~H. and {Goguen}, J. and {Johnson}, T.~V.},
        title = "{Voyager Disk-Integrated Photometry of Triton}",
      journal = {Science},
         year = 1990,
        month = oct,
       volume = {250},
       number = {4979},
        pages = {419-421},
          doi = {10.1126/science.250.4979.419},
       adsurl = {https://ui.adsabs.harvard.edu/abs/1990Sci...250..419H}
}

@ARTICLE{Sromovsky2021,
       author = {{Sromovsky}, L.~A. and {Baines}, K.~H. and {Fry}, P.~M.},
        title = "{Evolution of Saturn's north polar color and cloud structure between 2012 and 2017 inferred from Cassini VIMS and ISS observations}",
      journal = {\icarus},
         year = 2021,
        month = jul,
       volume = {362},
          eid = {114409},
        pages = {114409},
          doi = {10.1016/j.icarus.2021.114409},
archivePrefix = {arXiv},
       eprint = {2108.00322},
 primaryClass = {astro-ph.EP},
       adsurl = {https://ui.adsabs.harvard.edu/abs/2021Icar..36214409S}
}

@ARTICLE{Pogorelyuk2022,
       author = {{Pogorelyuk}, Leonid and {Fitzgerald}, Riley and {Vlahakis}, Sophia and {Morgan}, Rhonda and {Cahoy}, Kerri},
        title = "{Deconfusing Detections in Directly Imaged Multiplanet Systems}",
      journal = {\apj},
         year = 2022,
        month = oct,
       volume = {937},
       number = {2},
          eid = {66},
        pages = {66},
          doi = {10.3847/1538-4357/ac8d56},
       adsurl = {https://ui.adsabs.harvard.edu/abs/2022ApJ...937...66P}
}

@ARTICLE{Hasler2025,
       author = {{Hasler}, Samantha N. and {Pogorelyuk}, Leonid and {Fitzgerald}, Riley and {Cahoy}, Kerri and {Morgan}, Rhonda},
        title = "{Leveraging Photometry for Deconfusion of Directly Imaged Multi-Planet Systems}",
      journal = {arXiv e-prints},
         year = 2025,
        month = oct,
          eid = {arXiv:2510.09841},
        pages = {arXiv:2510.09841},
          doi = {10.48550/arXiv.2510.09841},
archivePrefix = {arXiv},
       eprint = {2510.09841},
 primaryClass = {astro-ph.EP},
       adsurl = {https://ui.adsabs.harvard.edu/abs/2025arXiv251009841H}
}

@article{Bennett_2014,
   title={MOA-2011-BLG-262Lb: A SUB-EARTH-MASS MOON ORBITING A GAS GIANT PRIMARY OR A HIGH VELOCITY PLANETARY SYSTEM IN THE GALACTIC BULGE},
   volume={785},
   ISSN={1538-4357},
   url={http://dx.doi.org/10.1088/0004-637X/785/2/155},
   DOI={10.1088/0004-637x/785/2/155},
   number={2},
   journal={The Astrophysical Journal},
   publisher={American Astronomical Society},
   author={Bennett, D. P. and Batista, V. and Bond, I. A. and Bennett, C. S. and Suzuki, D. and Beaulieu, J.-P. and Udalski, A. and Donatowicz, J. and Bozza, V. and Abe, F. and et al.},
   year={2014},
   month={Apr},
   pages={155}
}

@ARTICLE{5725236,
  author={van der Walt, Stefan and Colbert, S. Chris and Varoquaux, Gael},
  journal={Computing in Science   Engineering}, 
  title={The NumPy Array: A Structure for Efficient Numerical Computation}, 
  year={2011},
  volume={13},
  number={2},
  pages={22-30},
  doi={10.1109/MCSE.2011.37}}

@ARTICLE{2013ApJ...765..131K,
       author = {{Kopparapu}, Ravi Kumar and {Ramirez}, Ramses and {Kasting}, James F. and {Eymet}, Vincent and {Robinson}, Tyler D. and {Mahadevan}, Suvrath and {Terrien}, Ryan C. and {Domagal-Goldman}, Shawn and {Meadows}, Victoria and {Deshpande}, Rohit},
        title = "{Habitable Zones around Main-sequence Stars: New Estimates}",
      journal = {\apj},
         year = 2013,
        month = mar,
       volume = {765},
       number = {2},
          eid = {131},
        pages = {131},
          doi = {10.1088/0004-637X/765/2/131},
archivePrefix = {arXiv},
       eprint = {1301.6674},
 primaryClass = {astro-ph.EP},
       adsurl = {https://ui.adsabs.harvard.edu/abs/2013ApJ...765..131K}
}

@ARTICLE{1998Sci...281.1342H,
       author = {{Hoffman}, Paul F. and {Kaufman}, Alan J. and {Halverson}, Galen P. and {Schrag}, Daniel P.},
        title = "{A Neoproterozoic Snowball Earth}",
      journal = {Science},
         year = 1998,
        month = aug,
       volume = {281},
        pages = {1342},
          doi = {10.1126/science.281.5381.1342},
       adsurl = {https://ui.adsabs.harvard.edu/abs/1998Sci...281.1342H}
}

@ARTICLE{2002TeNov..14..129H,
       author = {{Hoffman}, Paul F. and {Schrag}, Daniel P.},
        title = "{The snowball Earth hypothesis: testing the limits of global change}",
      journal = {Terra Nova},
         year = 2002,
        month = jun,
       volume = {14},
       number = {3},
        pages = {129-155},
          doi = {10.1046/j.1365-3121.2002.00408.x},
       adsurl = {https://ui.adsabs.harvard.edu/abs/2002TeNov..14..129H}
}

@ARTICLE{2005GSAB..117.1181H,
       author = {{Halverson}, Galen P. and {Hoffman}, Paul F. and {Schrag}, Daniel P. and {Maloof}, Adam C. and {Rice}, A. Hugh N.},
        title = "{Toward a Neoproterozoic composite carbon-isotope record}",
      journal = {Geological Society of America Bulletin},
         year = 2005,
        month = jan,
       volume = {117},
       number = {9},
        pages = {1181},
          doi = {10.1130/B25630.1},
       adsurl = {https://ui.adsabs.harvard.edu/abs/2005GSAB..117.1181H}
}

@ARTICLE{2010Sci...327.1241M,
       author = {{Macdonald}, Francis A. and {Schmitz}, Mark D. and {Crowley}, James L. and {Roots}, Charles F. and {Jones}, David S. and {Maloof}, Adam C. and {Strauss}, Justin V. and {Cohen}, Phoebe A. and {Johnston}, David T. and {Schrag}, Daniel P.},
        title = "{Calibrating the Cryogenian}",
      journal = {Science},
         year = 2010,
        month = mar,
       volume = {327},
       number = {5970},
        pages = {1241},
          doi = {10.1126/science.1183325},
       adsurl = {https://ui.adsabs.harvard.edu/abs/2010Sci...327.1241M}
}

@BOOK{2010ppc..book.....P,
       author = {{Pierrehumbert}, Raymond T.},
        title = "{Principles of Planetary Climate}",
         year = 2010,
         publisher = {Cambridge University Press}, 
       adsurl = {https://ui.adsabs.harvard.edu/abs/2010ppc..book.....P}
}

@ARTICLE{1961JGR....66.3033W,
       author = {{Watson}, K. and {Murray}, B.~C. and {Brown}, H.},
        title = "{The behavior of volatiles on the lunar surface}",
      journal = {\jgr},
         year = 1961,
        month = sep,
       volume = {66},
       number = {9},
        pages = {3033-3045},
          doi = {10.1029/JZ066i009p03033},
       adsurl = {https://ui.adsabs.harvard.edu/abs/1961JGR....66.3033W}
}

@ARTICLE{1999Icar..141..179V,
       author = {{Vasavada}, Ashwin R. and {Paige}, David A. and {Wood}, Stephen E.},
        title = "{Near-Surface Temperatures on Mercury and the Moon and the Stability of Polar Ice Deposits}",
      journal = {\icarus},
         year = 1999,
        month = oct,
       volume = {141},
       number = {2},
        pages = {179-193},
          doi = {10.1006/icar.1999.6175},
       adsurl = {https://ui.adsabs.harvard.edu/abs/1999Icar..141..179V}
}

@ARTICLE{2015Icar..255...58H,
       author = {{Hayne}, Paul O. and {Hendrix}, Amanda and {Sefton-Nash}, Elliot and {Siegler}, Matthew A. and {Lucey}, Paul G. and {Retherford}, Kurt D. and {Williams}, Jean-Pierre and {Greenhagen}, Benjamin T. and {Paige}, David A.},
        title = "{Evidence for exposed water ice in the Moon's south polar regions from Lunar Reconnaissance Orbiter ultraviolet albedo and temperature measurements}",
      journal = {\icarus},
         year = 2015,
        month = jul,
       volume = {255},
        pages = {58-69},
          doi = {10.1016/j.icarus.2015.03.032},
       adsurl = {https://ui.adsabs.harvard.edu/abs/2015Icar..255...58H}
}

@ARTICLE{2005JGRD..110.1111P,
       author = {{Pierrehumbert}, R.~T.},
        title = "{Climate dynamics of a hard snowball Earth}",
      journal = {Journal of Geophysical Research (Atmospheres)},
         year = 2005,
        month = jan,
       volume = {110},
       number = {D1},
          eid = {D01111},
        pages = {D01111},
          doi = {10.1029/2004JD005162},
       adsurl = {https://ui.adsabs.harvard.edu/abs/2005JGRD..110.1111P}
}

@ARTICLE{2004A&A...420.1153A,
       author = {{Arnold}, L. and {Schneider}, J.},
        title = "{The detectability of extrasolar planet surroundings. I. Reflected-light photometry of unresolved rings}",
      journal = {\aap},
         year = 2004,
        month = jun,
       volume = {420},
        pages = {1153-1162},
          doi = {10.1051/0004-6361:20035720},
archivePrefix = {arXiv},
       eprint = {astro-ph/0403330},
 primaryClass = {astro-ph},
       adsurl = {https://ui.adsabs.harvard.edu/abs/2004A&A...420.1153A}
}

@BOOK{2021pdaa.book.....N,
       author = {{{National Academies of Sciences}, Engineering, and Medicine}},
        title = "{Pathways to Discovery in Astronomy and Astrophysics for the 2020s}",
         year = 2021,
         publisher = "National Academies Press",
          doi = {10.17226/26141},
       adsurl = {https://ui.adsabs.harvard.edu/abs/2021pdaa.book.....N}
}

@article{madden_catalog_2018,
	title = {A {Catalog} of {Spectra}, {Albedos}, and {Colors} of {Solar} {System} {Bodies} for {Exoplanet} {Comparison}},
	volume = {18},
	issn = {1531-1074},
	url = {https://www.liebertpub.com/doi/10.1089/ast.2017.1763},
	doi = {10.1089/ast.2017.1763},
	number = {12},
	urldate = {2025-07-13},
	journal = {Astrobiology},
	author = {Madden, J.H. and Kaltenegger, Lisa},
	month = dec,
	year = {2018},
	note = {Publisher: Mary Ann Liebert, Inc., publishers},
	pages = {1559--1573},
}

@article{lundock_2009,
	title = {Tohoku-{Hiroshima}-{Nagoya} planetary spectra library: a method for characterizing planets in the visible to near infrared},
	volume = {507},
	copyright = {© ESO, 2009},
	issn = {0004-6361, 1432-0746},
	shorttitle = {Tohoku-{Hiroshima}-{Nagoya} planetary spectra library},
	url = {https://www.aanda.org/articles/aa/abs/2009/45/aa12581-09/aa12581-09.html},
	doi = {10.1051/0004-6361/200912581},
	language = {en},
	number = {3},
	urldate = {2025-07-13},
	journal = {Astronomy \& Astrophysics},
	author = {Lundock, R. and Ichikawa, T. and Okita, H. and Kurita, K. and Kawabata, K. S. and Uemura, M. and Yamashita, T. and Ohsugi, T. and Sato, S. and Kino, M.},
	month = dec,
	year = {2009},
	note = {Number: 3
Publisher: EDP Sciences},
	pages = {1649--1658},
}

@article{holler_surface_2016,
	title = {On the surface composition of {Triton}’s southern latitudes},
	volume = {267},
	issn = {0019-1035},
	url = {https://www.sciencedirect.com/science/article/pii/S0019103515005849},
	doi = {https://doi.org/10.1016/j.icarus.2015.12.027},
	journal = {Icarus},
	author = {Holler, B. J. and Young, L. A. and Grundy, W. M. and Olkin, C. B.},
	year = {2016},
	pages = {255--266},
}

@article{cruikshank_diameter_1979,
	title = {The diameter and reflectance of {Triton}},
	volume = {40},
	issn = {0019-1035},
	url = {https://www.sciencedirect.com/science/article/pii/0019103579900575},
	doi = {https://doi.org/10.1016/0019-1035(79)90057-5},
	number = {1},
	journal = {Icarus},
	author = {Cruikshank, D. P. and Stockton, Alan and Dyck, H. M. and Becklin, E. E. and Macy, W.},
	year = {1979},
	pages = {104--114},
}

@article{ciarniello_cassini-vims_2019,
	title = {Cassini-{VIMS} observations of {Saturn}’s main rings: {II}. {A} spectrophotometric study by means of {Monte} {Carlo} ray-tracing and {Hapke}’s theory},
	volume = {317},
	issn = {0019-1035},
	url = {https://www.sciencedirect.com/science/article/pii/S0019103518300307},
	doi = {https://doi.org/10.1016/j.icarus.2018.07.010},
	journal = {Icarus},
	author = {Ciarniello, M. and Filacchione, G. and D’Aversa, E. and Capaccioni, F. and Nicholson, P. D. and Cuzzi, J. N. and Clark, R. N. and Hedman, M. M. and Ore, C. M. Dalle and Cerroni, P. and Plainaki, C. and Spilker, L. J.},
	year = {2019},
	pages = {242--265},
}

@ARTICLE{2024ApJS..272...30H,
       author = {{Harada}, Caleb K. and {Dressing}, Courtney D. and {Kane}, Stephen R. and {Ardestani}, Bahareh Adami},
        title = "{Setting the Stage for the Search for Life with the Habitable Worlds Observatory: Properties of 164 Promising Planet-survey Targets}",
      journal = {\apjs},
         year = 2024,
        month = jun,
       volume = {272},
       number = {2},
          eid = {30},
        pages = {30},
          doi = {10.3847/1538-4365/ad3e81},
archivePrefix = {arXiv},
       eprint = {2401.03047},
 primaryClass = {astro-ph.EP},
       adsurl = {https://ui.adsabs.harvard.edu/abs/2024ApJS..272...30H}
}

@ARTICLE{2025ApJ...980...39L,
       author = {{Lu}, Tiger and {Li}, Gongjie and {Cassese}, Ben and {Lin}, D.~N.~C.},
        title = "{The Dynamical History of HIP-41378 f{\textemdash}Oblique Exorings Masquerading as a Puffy Planet}",
      journal = {\apj},
         year = 2025,
        month = feb,
       volume = {980},
       number = {1},
          eid = {39},
        pages = {39},
          doi = {10.3847/1538-4357/ada4b2},
archivePrefix = {arXiv},
       eprint = {2410.00641},
 primaryClass = {astro-ph.EP},
       adsurl = {https://ui.adsabs.harvard.edu/abs/2025ApJ...980...39L}
}

@ARTICLE{2021AJ....161..202L,
       author = {{Liang}, Yan and {Robnik}, Jakob and {Seljak}, Uro{\v{s}}},
        title = "{Kepler-90: Giant Transit-timing Variations Reveal a Super-puff}",
      journal = {\aj},
         year = 2021,
        month = apr,
       volume = {161},
       number = {4},
          eid = {202},
        pages = {202},
          doi = {10.3847/1538-3881/abe6a7},
archivePrefix = {arXiv},
       eprint = {2011.08515},
 primaryClass = {astro-ph.EP},
       adsurl = {https://ui.adsabs.harvard.edu/abs/2021AJ....161..202L}
}

@ARTICLE{Limbach2021,
       author = {{Limbach}, Mary Anne and {Vos}, Johanna M. and {Winn}, Joshua N. and {Heller}, Ren{\'e} and {Mason}, Jeffrey C. and {Schneider}, Adam C. and {Dai}, Fei},
        title = "{On the Detection of Exomoons Transiting Isolated Planetary-mass Objects}",
      journal = {\apjl},
         year = 2021,
        month = sep,
       volume = {918},
       number = {2},
          eid = {L25},
        pages = {L25},
          doi = {10.3847/2041-8213/ac1e2d},
archivePrefix = {arXiv},
       eprint = {2108.08323},
 primaryClass = {astro-ph.EP},
       adsurl = {https://ui.adsabs.harvard.edu/abs/2021ApJ...918L..25L}
}

@ARTICLE{2005Icar..175..111B,
       author = {{Bottke}, William F. and {Durda}, Daniel D. and {Nesvorn{\'y}}, David and {Jedicke}, Robert and {Morbidelli}, Alessandro and {Vokrouhlick{\'y}}, David and {Levison}, Hal},
        title = "{The fossilized size distribution of the main asteroid belt}",
      journal = {\icarus},
         year = 2005,
        month = may,
       volume = {175},
       number = {1},
        pages = {111-140},
          doi = {10.1016/j.icarus.2004.10.026},
       adsurl = {https://ui.adsabs.harvard.edu/abs/2005Icar..175..111B}
}

@ARTICLE{1989Sci...246.1422S,
       author = {{Smith}, B.~A. and {Soderblom}, L.~A. and {Banfield}, D. and {Barnet}, C. and {Basilevksy}, A.~T. and {Beebe}, R.~F. and {Bollinger}, K. and {Boyce}, J.~M. and {Brahic}, A. and {Briggs}, G.~A. and {Brown}, R.~H. and {Chyba}, C. and {Collins}, S.~A. and {Colvin}, T. and {Cook}, A.~F. and {Crisp}, D. and {Croft}, S.~K. and {Cruikshank}, D. and {Cuzzi}, J.~N. and {Danielson}, G.~E. and {Davies}, M.~E. and {de Jong}, E. and {Dones}, L. and {Godfrey}, D. and {Goguen}, J. and {Grenier}, I. and {Haemmerle}, V.~R. and {Hammel}, H. and {Hansen}, C.~J. and {Helfenstein}, C.~P. and {Howell}, C. and {Hunt}, G.~E. and {Ingersoll}, A.~P. and {Johnson}, T.~V. and {Kargel}, J. and {Kirk}, R. and {Kuehn}, D.~I. and {Limaye}, S. and {Masursky}, H. and {McEwen}, A. and {Morrison}, D. and {Owen}, T. and {Owen}, W. and {Pollack}, J.~B. and {Porco}, C.~C. and {Rages}, K. and {Rogers}, P. and {Rudy}, D. and {Sagan}, C. and {Schwartz}, J. and {Shoemaker}, E.~M. and {Showalter}, M. and {Sicardy}, B. and {Simonelli}, D. and {Spencer}, J. and {Sromovsky}, L.~A. and {Stoker}, C. and {Strom}, R.~G. and {Suomi}, V.~E. and {Synott}, S.~P. and {Terrile}, R.~J. and {Thomas}, P. and {Thompson}, W.~R. and {Verbiscer}, A. and {Veverka}, J.},
        title = "{Voyager 2 at Neptune: Imaging Science Results}",
      journal = {Science},
         year = 1989,
        month = dec,
       volume = {246},
       number = {4936},
        pages = {1422-1449},
          doi = {10.1126/science.246.4936.1422},
       adsurl = {https://ui.adsabs.harvard.edu/abs/1989Sci...246.1422S}
}

@ARTICLE{1986Natur.319..636H,
       author = {{Hubbard}, W.~B. and {Brahic}, A. and {Sicardy}, B. and {Elicer}, L. -R. and {Roques}, F. and {Vilas}, F.},
        title = "{Occultation detection of a neptunian ring-like arc}",
      journal = {\nat},
         year = 1986,
        month = feb,
       volume = {319},
       number = {6055},
        pages = {636-640},
          doi = {10.1038/319636a0},
       adsurl = {https://ui.adsabs.harvard.edu/abs/1986Natur.319..636H}
}

@ARTICLE{1977Natur.267..328E,
       author = {{Elliot}, J.~L. and {Dunham}, E. and {Mink}, D.},
        title = "{The rings of Uranus}",
      journal = {\nat},
         year = 1977,
        month = may,
       volume = {267},
       number = {5609},
        pages = {328-330},
          doi = {10.1038/267328a0},
       adsurl = {https://ui.adsabs.harvard.edu/abs/1977Natur.267..328E}
}

@ARTICLE{2010Natur.468..943C,
       author = {{Canup}, Robin M.},
        title = "{Origin of Saturn's rings and inner moons by mass removal from a lost Titan-sized satellite}",
      journal = {\nat},
         year = 2010,
        month = dec,
       volume = {468},
       number = {7326},
        pages = {943-946},
          doi = {10.1038/nature09661},
       adsurl = {https://ui.adsabs.harvard.edu/abs/2010Natur.468..943C}
}

@ARTICLE{2022Sci...377.1285W,
       author = {{Wisdom}, Jack and {Dbouk}, Rola and {Militzer}, Burkhard and {Hubbard}, William B. and {Nimmo}, Francis and {Downey}, Brynna G. and {French}, Richard G.},
        title = "{Loss of a satellite could explain Saturn{\textquoteright}s obliquity and young rings}",
      journal = {Science},
         year = 2022,
        month = sep,
       volume = {377},
       number = {6612},
        pages = {1285-1289},
          doi = {10.1126/science.abn1234},
       adsurl = {https://ui.adsabs.harvard.edu/abs/2022Sci...377.1285W}
}

@INCOLLECTION{2018haex.bookE.158C,
       author = {{Cabrera}, Juan and {Jim{\'e}nez}, Mar{\'\i}a Fern{\'a}ndez and {Mu{\~n}oz}, Antonio Garc{\'\i}a and {Schneider}, Jean},
        title = "{Special Cases: Moons, Rings, Comets, and Trojans}",
    booktitle = {Handbook of Exoplanets},
         year = 2018,
         publisher = {Springer Nature},
       editor = {{Deeg}, Hans J. and {Belmonte}, Juan Antonio},
          eid = {158},
        pages = {158},
          doi = {10.1007/978-3-319-55333-7_158},
       adsurl = {https://ui.adsabs.harvard.edu/abs/2018haex.bookE.158C}
}

@ARTICLE{2022ApJ...934L..20B,
       author = {{Bae}, Jaehan and {Teague}, Richard and {Andrews}, Sean M. and {Benisty}, Myriam and {Facchini}, Stefano and {Galloway-Sprietsma}, Maria and {Loomis}, Ryan A. and {Aikawa}, Yuri and {Alarc{\'o}n}, Felipe and {Bergin}, Edwin and {Bergner}, Jennifer B. and {Booth}, Alice S. and {Cataldi}, Gianni and {Cleeves}, L. Ilsedore and {Czekala}, Ian and {Guzm{\'a}n}, Viviana V. and {Huang}, Jane and {Ilee}, John D. and {Kurtovic}, Nicolas T. and {Law}, Charles J. and {Le Gal}, Romane and {Liu}, Yao and {Long}, Feng and {M{\'e}nard}, Fran{\c{c}}ois and {{\"O}berg}, Karin I. and {P{\'e}rez}, Laura M. and {Qi}, Chunhua and {Schwarz}, Kamber R. and {Sierra}, Anibal and {Walsh}, Catherine and {Wilner}, David J. and {Zhang}, Ke},
        title = "{Molecules with ALMA at Planet-forming Scales (MAPS): A Circumplanetary Disk Candidate in Molecular-line Emission in the AS 209 Disk}",
      journal = {\apjl},
         year = 2022,
        month = aug,
       volume = {934},
       number = {2},
          eid = {L20},
        pages = {L20},
          doi = {10.3847/2041-8213/ac7fa3},
archivePrefix = {arXiv},
       eprint = {2207.05923},
 primaryClass = {astro-ph.EP},
       adsurl = {https://ui.adsabs.harvard.edu/abs/2022ApJ...934L..20B}
}

@ARTICLE{2022ApJ...940L..30O,
       author = {{Ohno}, Kazumasa and {Thao}, Pa Chia and {Mann}, Andrew W. and {Fortney}, Jonathan J.},
        title = "{A Circumplanetary Dust Ring May Explain the Extreme Spectral Slope of the 10 Myr Young Exoplanet K2-33b}",
      journal = {\apjl},
         year = 2022,
        month = dec,
       volume = {940},
       number = {2},
          eid = {L30},
        pages = {L30},
          doi = {10.3847/2041-8213/ac9f3f},
archivePrefix = {arXiv},
       eprint = {2211.07706},
 primaryClass = {astro-ph.EP},
       adsurl = {https://ui.adsabs.harvard.edu/abs/2022ApJ...940L..30O}
}

@ARTICLE{2021ApJ...916L...2B,
       author = {{Benisty}, Myriam and {Bae}, Jaehan and {Facchini}, Stefano and {Keppler}, Miriam and {Teague}, Richard and {Isella}, Andrea and {Kurtovic}, Nicolas T. and {P{\'e}rez}, Laura M. and {Sierra}, Anibal and {Andrews}, Sean M. and {Carpenter}, John and {Czekala}, Ian and {Dominik}, Carsten and {Henning}, Thomas and {Menard}, Francois and {Pinilla}, Paola and {Zurlo}, Alice},
        title = "{A Circumplanetary Disk around PDS70c}",
      journal = {\apjl},
         year = 2021,
        month = jul,
       volume = {916},
       number = {1},
          eid = {L2},
        pages = {L2},
          doi = {10.3847/2041-8213/ac0f83},
archivePrefix = {arXiv},
       eprint = {2108.07123},
 primaryClass = {astro-ph.EP},
       adsurl = {https://ui.adsabs.harvard.edu/abs/2021ApJ...916L...2B}
}

@ARTICLE{2015NatGe...8..686H,
       author = {{Hyodo}, Ryuki and {Ohtsuki}, Keiji},
        title = "{Saturn's F ring and shepherd satellites a natural outcome of satellite system formation}",
      journal = {Nature Geoscience},
         year = 2015,
        month = sep,
       volume = {8},
       number = {9},
        pages = {686-689},
          doi = {10.1038/ngeo2508},
       adsurl = {https://ui.adsabs.harvard.edu/abs/2015NatGe...8..686H}
}

@ARTICLE{2015ApJ...800..126K,
       author = {{Kenworthy}, M.~A. and {Mamajek}, E.~E.},
        title = "{Modeling Giant Extrasolar Ring Systems in Eclipse and the Case of J1407b: Sculpting by Exomoons?}",
      journal = {\apj},
         year = 2015,
        month = feb,
       volume = {800},
       number = {2},
          eid = {126},
        pages = {126},
          doi = {10.1088/0004-637X/800/2/126},
archivePrefix = {arXiv},
       eprint = {1501.05652},
 primaryClass = {astro-ph.SR},
       adsurl = {https://ui.adsabs.harvard.edu/abs/2015ApJ...800..126K}
}

@ARTICLE{1979Sci...204..951S,
       author = {{Smith}, Bradford A. and {Soderblom}, Laurence A. and {Johnson}, Torrence V. and {Ingersoll}, Andrew P. and {Collins}, Stewart A. and {Shoemaker}, Eugene M. and {Hunt}, G.~E. and {Masursky}, Harold and {Carr}, Michael H. and {Davies}, Merton E. and {Cook}, Allan F., II and {Boyce}, Joseph and {Danielson}, G. Edward and {Owen}, Tobias and {Sagan}, Carl and {Beebe}, Reta F. and {Veverka}, Joseph and {Strom}, Robert G. and {McCauley}, John F. and {Morrison}, David and {Briggs}, Geoffrey A. and {Suomi}, Verner E.},
        title = "{The Jupiter System Through the Eyes of Voyager 1}",
      journal = {Science},
         year = 1979,
        month = jun,
       volume = {204},
       number = {4396},
        pages = {951-957},
          doi = {10.1126/science.204.4396.951},
       adsurl = {https://ui.adsabs.harvard.edu/abs/1979Sci...204..951S}
}

@ARTICLE{2011ApJ...741...51R,
       author = {{Robinson}, Tyler D.},
        title = "{Modeling the Infrared Spectrum of the Earth-Moon System: Implications for the Detection and Characterization of Earthlike Extrasolar Planets and Their Moonlike Companions}",
      journal = {\apj},
         year = 2011,
        month = nov,
       volume = {741},
       number = {1},
          eid = {51},
        pages = {51},
          doi = {10.1088/0004-637X/741/1/51},
archivePrefix = {arXiv},
       eprint = {1110.3744},
 primaryClass = {astro-ph.EP},
       adsurl = {https://ui.adsabs.harvard.edu/abs/2011ApJ...741...51R}
}

@ARTICLE{2014PNAS..111.6871R,
       author = {{Rein}, H. and {Fujii}, Y. and {Spiegel}, D.~S.},
        title = "{Some inconvenient truths about biosignatures involving two chemical species on Earth-like exoplanets}",
      journal = {Proceedings of the National Academy of Science},
         year = 2014,
        month = may,
       volume = {111},
       number = {19},
        pages = {6871-6875},
          doi = {10.1073/pnas.1401816111},
archivePrefix = {arXiv},
       eprint = {1404.6531},
 primaryClass = {astro-ph.EP},
       adsurl = {https://ui.adsabs.harvard.edu/abs/2014PNAS..111.6871R}
}

@ARTICLE{2004AsBio...4..400W,
       author = {{Williams}, D.~M. and {Knacke}, R.~F.},
        title = "{Looking for Planetary Moons in the Spectra of Distant Jupiters}",
      journal = {Astrobiology},
         year = 2004,
        month = sep,
       volume = {4},
       number = {3},
        pages = {400-403},
          doi = {10.1089/ast.2004.4.400},
       adsurl = {https://ui.adsabs.harvard.edu/abs/2004AsBio...4..400W}
}

@ARTICLE{2020AJ....159..131P,
       author = {{Piro}, Anthony L. and {Vissapragada}, Shreyas},
        title = "{Exploring Whether Super-puffs can be Explained as Ringed Exoplanets}",
      journal = {\aj},
         year = 2020,
        month = apr,
       volume = {159},
       number = {4},
          eid = {131},
        pages = {131},
          doi = {10.3847/1538-3881/ab7192},
archivePrefix = {arXiv},
       eprint = {1911.09673},
 primaryClass = {astro-ph.EP},
       adsurl = {https://ui.adsabs.harvard.edu/abs/2020AJ....159..131P}
}

@ARTICLE{2003A&A...412..305P,
       author = {{Poulet}, F. and {Cruikshank}, D.~P. and {Cuzzi}, J.~N. and {Roush}, T.~L. and {French}, R.~G.},
        title = "{Compositions of Saturn's rings A, B, and C from high resolution near-infrared spectroscopic observations}",
      journal = {\aap},
         year = 2003,
        month = dec,
       volume = {412},
        pages = {305-316},
          doi = {10.1051/0004-6361:20031123},
       adsurl = {https://ui.adsabs.harvard.edu/abs/2003A&A...412..305P}
}

@ARTICLE{2023Natur.614..239M,
       author = {{Morgado}, B.~E. and {Sicardy}, B. and {Braga-Ribas}, F. and {Ortiz}, J.~L. and {Salo}, H. and {Vachier}, F. and {Desmars}, J. and {Pereira}, C.~L. and {Santos-Sanz}, P. and {Sfair}, R. and {de Santana}, T. and {Assafin}, M. and {Vieira-Martins}, R. and {Gomes-J{\'u}nior}, A.~R. and {Margoti}, G. and {Dhillon}, V.~S. and {Fern{\'a}ndez-Valenzuela}, E. and {Broughton}, J. and {Bradshaw}, J. and {Langersek}, R. and {Benedetti-Rossi}, G. and {Souami}, D. and {Holler}, B.~J. and {Kretlow}, M. and {Boufleur}, R.~C. and {Camargo}, J.~I.~B. and {Duffard}, R. and {Beisker}, W. and {Morales}, N. and {Lecacheux}, J. and {Rommel}, F.~L. and {Herald}, D. and {Benz}, W. and {Jehin}, E. and {Jankowsky}, F. and {Marsh}, T.~R. and {Littlefair}, S.~P. and {Bruno}, G. and {Pagano}, I. and {Brandeker}, A. and {Collier-Cameron}, A. and {Flor{\'e}n}, H.~G. and {Hara}, N. and {Olofsson}, G. and {Wilson}, T.~G. and {Benkhaldoun}, Z. and {Busuttil}, R. and {Burdanov}, A. and {Ferrais}, M. and {Gault}, D. and {Gillon}, M. and {Hanna}, W. and {Kerr}, S. and {Kolb}, U. and {Nosworthy}, P. and {Sebastian}, D. and {Snodgrass}, C. and {Teng}, J.~P. and {de Wit}, J.},
        title = "{A dense ring of the trans-Neptunian object Quaoar outside its Roche limit}",
      journal = {\nat},
         year = 2023,
        month = feb,
       volume = {614},
       number = {7947},
        pages = {239-243},
          doi = {10.1038/s41586-022-05629-6},
       adsurl = {https://ui.adsabs.harvard.edu/abs/2023Natur.614..239M}
}

@ARTICLE{2017Natur.550..219O,
       author = {{Ortiz}, J.~L. and {Santos-Sanz}, P. and {Sicardy}, B. and {Benedetti-Rossi}, G. and {B{\'e}rard}, D. and {Morales}, N. and {Duffard}, R. and {Braga-Ribas}, F. and {Hopp}, U. and {Ries}, C. and {Nascimbeni}, V. and {Marzari}, F. and {Granata}, V. and {P{\'a}l}, A. and {Kiss}, C. and {Pribulla}, T. and {Kom{\v{z}}{\'\i}k}, R. and {Hornoch}, K. and {Pravec}, P. and {Bacci}, P. and {Maestripieri}, M. and {Nerli}, L. and {Mazzei}, L. and {Bachini}, M. and {Martinelli}, F. and {Succi}, G. and {Ciabattari}, F. and {Mikuz}, H. and {Carbognani}, A. and {Gaehrken}, B. and {Mottola}, S. and {Hellmich}, S. and {Rommel}, F.~L. and {Fern{\'a}ndez-Valenzuela}, E. and {Campo Bagatin}, A. and {Cikota}, S. and {Cikota}, A. and {Lecacheux}, J. and {Vieira-Martins}, R. and {Camargo}, J.~I.~B. and {Assafin}, M. and {Colas}, F. and {Behrend}, R. and {Desmars}, J. and {Meza}, E. and {Alvarez-Candal}, A. and {Beisker}, W. and {Gomes-Junior}, A.~R. and {Morgado}, B.~E. and {Roques}, F. and {Vachier}, F. and {Berthier}, J. and {Mueller}, T.~G. and {Madiedo}, J.~M. and {Unsalan}, O. and {Sonbas}, E. and {Karaman}, N. and {Erece}, O. and {Koseoglu}, D.~T. and {Ozisik}, T. and {Kalkan}, S. and {Guney}, Y. and {Niaei}, M.~S. and {Satir}, O. and {Yesilyaprak}, C. and {Puskullu}, C. and {Kabas}, A. and {Demircan}, O. and {Alikakos}, J. and {Charmandaris}, V. and {Leto}, G. and {Ohlert}, J. and {Christille}, J.~M. and {Szak{\'a}ts}, R. and {Tak{\'a}csn{\'e} Farkas}, A. and {Varga-Vereb{\'e}lyi}, E. and {Marton}, G. and {Marciniak}, A. and {Bartczak}, P. and {Santana-Ros}, T. and {Butkiewicz-B{k{a}}k}, M. and {Dudzi{\'n}ski}, G. and {Al{\'\i}-Lagoa}, V. and {Gazeas}, K. and {Tzouganatos}, L. and {Paschalis}, N. and {Tsamis}, V. and {S{\'a}nchez-Lavega}, A. and {P{\'e}rez-Hoyos}, S. and {Hueso}, R. and {Guirado}, J.~C. and {Peris}, V. and {Iglesias-Marzoa}, R.},
        title = "{The size, shape, density and ring of the dwarf planet Haumea from a stellar occultation}",
      journal = {\nat},
         year = 2017,
        month = oct,
       volume = {550},
       number = {7675},
        pages = {219-223},
          doi = {10.1038/nature24051},
archivePrefix = {arXiv},
       eprint = {2006.03113},
 primaryClass = {astro-ph.EP},
       adsurl = {https://ui.adsabs.harvard.edu/abs/2017Natur.550..219O}
}

@ARTICLE{2006A&A...450..395S,
       author = {{Szab{\'o}}, Gy. M. and {Szatm{\'a}ry}, K. and {Div{\'e}ki}, Zs. and {Simon}, A.},
        title = "{Possibility of a photometric detection of ``exomoons''}",
      journal = {\aap},
         year = 2006,
        month = apr,
       volume = {450},
       number = {1},
        pages = {395-398},
          doi = {10.1051/0004-6361:20054555},
archivePrefix = {arXiv},
       eprint = {astro-ph/0601186},
 primaryClass = {astro-ph},
       adsurl = {https://ui.adsabs.harvard.edu/abs/2006A&A...450..395S}
}

@ARTICLE{2007A&A...470..727S,
       author = {{Simon}, A. and {Szatm{\'a}ry}, K. and {Szab{\'o}}, Gy. M.},
        title = "{Determination of the size, mass, and density of ``exomoons'' from photometric transit timing variations}",
      journal = {\aap},
         year = 2007,
        month = aug,
       volume = {470},
       number = {2},
        pages = {727-731},
          doi = {10.1051/0004-6361:20066560},
archivePrefix = {arXiv},
       eprint = {0705.1046},
 primaryClass = {astro-ph},
       adsurl = {https://ui.adsabs.harvard.edu/abs/2007A&A...470..727S}
}

@ARTICLE{2018AJ....156..184V,
       author = {{Vanderburg}, Andrew and {Rappaport}, Saul A. and {Mayo}, Andrew W.},
        title = "{Detecting Exomoons via Doppler Monitoring of Directly Imaged Exoplanets}",
      journal = {\aj},
         year = 2018,
        month = nov,
       volume = {156},
       number = {5},
          eid = {184},
        pages = {184},
          doi = {10.3847/1538-3881/aae0fc},
archivePrefix = {arXiv},
       eprint = {1805.01903},
 primaryClass = {astro-ph.EP},
       adsurl = {https://ui.adsabs.harvard.edu/abs/2018AJ....156..184V}
}

@ARTICLE{2025arXiv250915304W,
       author = {{Winterhalder}, T.~O. and {M{\'e}rand}, A. and {Kammerer}, J. and {Lacour}, S. and {Nowak}, M. and {Balmer}, W.~O. and {Bourdarot}, G. and {Eisenhauer}, F. and {Glindemann}, A. and {Grant}, S. and {Henning}, Th. and {Kervella}, P. and {Marleau}, G. -D. and {Pourr{\'e}}, N. and {Rickman}, E.},
        title = "{Astrometric exomoon detection by means of optical interferometry}",
      journal = {arXiv e-prints},
         year = 2025,
        month = sep,
          eid = {arXiv:2509.15304},
        pages = {arXiv:2509.15304},
          doi = {10.48550/arXiv.2509.15304},
archivePrefix = {arXiv},
       eprint = {2509.15304},
 primaryClass = {astro-ph.EP},
       adsurl = {https://ui.adsabs.harvard.edu/abs/2025arXiv250915304W}
}

@ARTICLE{2010A&A...520A..68L,
       author = {{Liebig}, C. and {Wambsganss}, J.},
        title = "{Detectability of extrasolar moons as gravitational microlenses}",
      journal = {\aap},
         year = 2010,
        month = sep,
       volume = {520},
          eid = {A68},
        pages = {A68},
          doi = {10.1051/0004-6361/200913844},
archivePrefix = {arXiv},
       eprint = {0912.2076},
 primaryClass = {astro-ph.EP},
       adsurl = {https://ui.adsabs.harvard.edu/abs/2010A&A...520A..68L}
}

@ARTICLE{2024AJ....168...54L,
       author = {{Limbach}, Mary Anne and {Vos}, Johanna M. and {Vanderburg}, Andrew and {Dai}, Fei},
        title = "{Occurrence Rates of Exosatellites Orbiting 3{\textendash}30 M $_{Jup}$ Hosts from 44 Spitzer Light Curves}",
      journal = {\aj},
         year = 2024,
        month = aug,
       volume = {168},
       number = {2},
          eid = {54},
        pages = {54},
          doi = {10.3847/1538-3881/ad4ed5},
archivePrefix = {arXiv},
       eprint = {2405.08116},
 primaryClass = {astro-ph.EP},
       adsurl = {https://ui.adsabs.harvard.edu/abs/2024AJ....168...54L}
}

@ARTICLE{2025arXiv251024575W,
       author = {{Wilson}, Mikayla J. and {Limbach}, Mary Anne and {Skemer}, Andrew J. and {Vos}, Johanna M. and {Miles}, Brittany E. and {Rowland}, Melanie J. and {Vanderburg}, Andrew and {Schneider}, Adam C. and {Morley}, Caroline and {Kotten}, Brooke and {Householder}, Andrew and {Lupu}, Roxana and {Mang}, James and {Freedman}, Richard},
        title = "{A Deep Search for Exomoons Around WISE 0855 With JWST}",
      journal = {arXiv e-prints},
         year = 2025,
        month = oct,
          eid = {arXiv:2510.24575},
        pages = {arXiv:2510.24575},
archivePrefix = {arXiv},
       eprint = {2510.24575},
 primaryClass = {astro-ph.EP},
       adsurl = {https://ui.adsabs.harvard.edu/abs/2025arXiv251024575W}
}

@ARTICLE{2023AJ....165..113R,
       author = {{Ruffio}, Jean-Baptiste and {Horstman}, Katelyn and {Mawet}, Dimitri and {Rosenthal}, Lee J. and {Batygin}, Konstantin and {Wang}, Jason J. and {Millar-Blanchaer}, Maxwell and {Wang}, Ji and {Fulton}, Benjamin J. and {Konopacky}, Quinn M. and {Agrawal}, Shubh and {Hirsch}, Lea A. and {Howard}, Andrew W. and {Blunt}, Sarah and {Nielsen}, Eric and {Baker}, Ashley and {Bartos}, Randall and {Bond}, Charlotte Z. and {Calvin}, Benjamin and {Cetre}, Sylvain and {Delorme}, Jacques-Robert and {Doppmann}, Greg and {Echeverri}, Daniel and {Finnerty}, Luke and {Fitzgerald}, Michael P. and {Jovanovic}, Nemanja and {L{\'o}pez}, Ronald and {Martin}, Emily C. and {Morris}, Evan and {Pezzato}, Jacklyn and {Ruane}, Garreth and {Sappey}, Ben and {Schofield}, Tobias and {Skemer}, Andrew and {Venenciano}, Taylor and {Wallace}, J. Kent and {Wallack}, Nicole L. and {Wizinowich}, Peter and {Xuan}, Jerry W.},
        title = "{Detecting Exomoons from Radial Velocity Measurements of Self-luminous Planets: Application to Observations of HR 7672 B and Future Prospects}",
      journal = {\aj},
         year = 2023,
        month = mar,
       volume = {165},
       number = {3},
          eid = {113},
        pages = {113},
          doi = {10.3847/1538-3881/acb34a},
archivePrefix = {arXiv},
       eprint = {2301.04206},
 primaryClass = {astro-ph.EP},
       adsurl = {https://ui.adsabs.harvard.edu/abs/2023AJ....165..113R}
}

@ARTICLE{2025Natur.643..938H,
       author = {{Hoch}, K.~K.~W. and {Rowland}, M. and {Petrus}, S. and {Nasedkin}, E. and {Ingebretsen}, C. and {Kammerer}, J. and {Perrin}, M. and {D'Orazi}, V. and {Balmer}, W.~O. and {Barman}, T. and {Bonnefoy}, M. and {Chauvin}, G. and {Chen}, C. and {De Rosa}, R.~J. and {Girard}, J. and {Gonzales}, E. and {Kenworthy}, M. and {Konopacky}, Q.~M. and {Macintosh}, B. and {Moran}, S.~E. and {Morley}, C.~V. and {Palma-Bifani}, P. and {Pueyo}, L. and {Ren}, B. and {Rickman}, E. and {Ruffio}, J.-B. and {Theissen}, C.~A. and {Ward-Duong}, K. and {Zhang}, Y.},
        title = "{Silicate clouds and a circumplanetary disk in the YSES-1 exoplanet system}",
      journal = {\nat},
         year = 2025,
        month = jul,
       volume = {643},
       number = {8073},
        pages = {938-942},
          doi = {10.1038/s41586-025-09174-w},
archivePrefix = {arXiv},
       eprint = {2507.18861},
 primaryClass = {astro-ph.EP},
       adsurl = {https://ui.adsabs.harvard.edu/abs/2025Natur.643..938H}
}

@article{Pearl_neptune_bond_albedo,
author = {Pearl, J. C. and Conrath, B. J.},
title = {The albedo, effective temperature, and energy balance of Neptune, as determined from Voyager data},
journal = {Journal of Geophysical Research: Space Physics},
volume = {96},
number = {S01},
pages = {18921-18930},
doi = {https://doi.org/10.1029/91JA01087},
url = {https://agupubs.onlinelibrary.wiley.com/doi/abs/10.1029/91JA01087},
eprint = {https://agupubs.onlinelibrary.wiley.com/doi/pdf/10.1029/91JA01087},
year = {1991}
}

@article{HANEL1983262,
title = {Albedo, internal heat flux, and energy balance of Saturn},
journal = {Icarus},
volume = {53},
number = {2},
pages = {262-285},
year = {1983},
issn = {0019-1035},
doi = {https://doi.org/10.1016/0019-1035(83)90147-1},
url = {https://www.sciencedirect.com/science/article/pii/0019103583901471},
author = {R.A. Hanel and B.J. Conrath and V.G. Kunde and J.C. Pearl and J.A. Pirraglia}
}

@article{PEARL199012,
title = {The albedo, effective temperature, and energy balance of Uranus, as determined from Voyager IRIS data},
journal = {Icarus},
volume = {84},
number = {1},
pages = {12-28},
year = {1990},
issn = {0019-1035},
doi = {https://doi.org/10.1016/0019-1035(90)90155-3},
url = {https://www.sciencedirect.com/science/article/pii/0019103590901553},
author = {J.C. Pearl and B.J. Conrath and R.A. Hanel and J.A. Pirraglia and A. Coustenis}
}

@Article{rs14061518,
AUTHOR = {Simon, Amy A. and Wong, Michael H. and Sromovsky, Lawrence A. and Fletcher, Leigh N. and Fry, Patrick M.},
TITLE = {Giant Planet Atmospheres: Dynamics and Variability from UV to Near-IR Hubble and Adaptive Optics Imaging},
JOURNAL = {Remote Sensing},
VOLUME = {14},
YEAR = {2022},
NUMBER = {6},
ARTICLE-NUMBER = {1518},
URL = {https://www.mdpi.com/2072-4292/14/6/1518},
ISSN = {2072-4292},
DOI = {10.3390/rs14061518}
}

@article{2000GL012580,
author = {Goode, P. R. and Qiu, J. and Yurchyshyn, V. and Hickey, J. and Chu, M.-C. and Kolbe, E. and Brown, C. T. and Koonin, S. E.},
title = {Earthshine observations of the Earth's reflectance},
journal = {Geophysical Research Letters},
volume = {28},
number = {9},
pages = {1671-1674},
doi = {https://doi.org/10.1029/2000GL012580},
url = {https://agupubs.onlinelibrary.wiley.com/doi/abs/10.1029/2000GL012580},
eprint = {https://agupubs.onlinelibrary.wiley.com/doi/pdf/10.1029/2000GL012580},
year = {2001}
}

@article{Coulter2022Jupiter,
	author = {Coulter, Daniel J. and Barnes, Jason W. and Fortney, Jonathan J.},
	journal = {The Astrophysical Journal Supplement Series},
	doi = {10.3847/1538-4365/ac886a},
	issn = {0067-0049},
	number = {1},
	year = {2022},
	month = {nov 1},
	pages = {15},
	publisher = {American Astronomical Society},
	title = {Jupiter and {Saturn} as {Spectral} {Analogs} for {Extrasolar} {Gas} {Giants} and {Brown} {Dwarfs}},
	url = {http://dx.doi.org/10.3847/1538-4365/ac886a},
	volume = {263},
}

@article{BAINES2005ATMOSPHERES,
	author = {Baines, K. H. and Drossart, P. and Momary, T. W. and Formisano, V. and Griffith, C. and Bellucci, G. and Bibring, J. P. and Brown, R. H. and Buratti, B. J. and Capaccioni, F. and Cerroni, P. and Clark, R. N. and Coradini, A. and Combes, M. and Cruikshank, D. P. and Jaumann, R. and Langevin, Y. and Matson, D. L. and Mccord, T. B. and Mennella, V. and Nelson, R. M. and Nicholson, P. D. and Sicardy, B. and Sotin, C.},
	journal = {Earth, Moon, and Planets},
	doi = {10.1007/s11038-005-9058-2},
	issn = {0167-9295},
	number = {3-4},
	year = {2005},
	month = {6},
	pages = {119--147},
	publisher = {{Springer Science and Business Media LLC}},
	title = {THE {ATMOSPHERES} {OF} {SATURN} {AND} {TITAN} {IN} {THE} {NEAR}-{INFRARED}: FIRST {RESULTS} {OF} {CASSINI}/{VIMS}},
	url = {http://dx.doi.org/10.1007/s11038-005-9058-2},
	volume = {96},
}

@article{Seidelmann2007Report,
	author = {Seidelmann, P. Kenneth and Archinal, B. A. and A\textquoteright{}hearn, M. F. and Conrad, A. and Consolmagno, G. J. and Hestroffer, D. and Hilton, J. L. and Krasinsky, G. A. and Neumann, G. and Oberst, J. and Stooke, P. and Tedesco, E. F. and Tholen, D. J. and Thomas, P. C. and Williams, I. P.},
	journal = {Celestial Mechanics and Dynamical Astronomy},
	doi = {10.1007/s10569-007-9072-y},
	issn = {0923-2958},
	number = {3},
	year = {2007},
	month = {7},
	pages = {155--180},
	publisher = {{Springer Science and Business Media LLC}},
	title = {Report of the {IAU}/{IAG} {Working} {Group} on cartographic coordinates and rotational elements: 2006},
	url = {http://dx.doi.org/10.1007/s10569-007-9072-y},
	volume = {98},
}

@misc{Murray2000Solar,
	author = {Murray, Carl D. and Dermott, Stanley F.},
	doi = {10.1017/cbo9781139174817},
	isbn = {9780521572958},
	year = {2000},
	month = {feb 13},
	title = {Solar {System} {Dynamics}},
	url = {http://dx.doi.org/10.1017/CBO9781139174817},
}

@article{Fletcher2010Neptune,
	author = {Fletcher, L. N. and Drossart, P. and Burgdorf, M. and Orton, G. S. and Encrenaz, T.},
	journal = {Astronomy and Astrophysics},
	doi = {10.1051/0004-6361/200913358},
	issn = {0004-6361},
	year = {2010},
	month = {5},
	pages = {A17},
	publisher = {EDP Sciences},
	title = {Neptune's atmospheric composition from {AKARI} infrared spectroscopy},
	url = {http://dx.doi.org/10.1051/0004-6361/200913358},
	volume = {514},
}

@article{Lazzoni_2022,
   title={Detectability of satellites around directly imaged exoplanets and brown dwarfs},
   volume={516},
   ISSN={1365-2966},
   url={http://dx.doi.org/10.1093/mnras/stac2081},
   DOI={10.1093/mnras/stac2081},
   number={1},
   journal={Monthly Notices of the Royal Astronomical Society},
   publisher={Oxford University Press (OUP)},
   author={Lazzoni, Cecilia and Desidera, Silvano and Gratton, Raffaele and Zurlo, Alice and Mesa, Dino and Ray, Shrishmoy},
   year={2022},
   month=aug, pages={391–409} }

@article{Agol2015CENTER,
	author = {Agol, Eric and Jansen, Tiffany and Lacy, Brianna and Robinson, Tyler D. and Meadows, Victoria},
	journal = {The Astrophysical Journal},
	doi = {10.1088/0004-637x/812/1/5},
	issn = {1538-4357},
	number = {1},
	year = {2015},
	month = {oct 2},
	pages = {5},
	publisher = {American Astronomical Society},
	title = {THE {CENTER} {OF} {LIGHT}: SPECTROASTROMETRIC {DETECTION} {OF} {EXOMOONS}},
	url = {http://dx.doi.org/10.1088/0004-637X/812/1/5},
	volume = {812},
}

@misc{hwo-tools,
  title        = {{hwo-tools}: Software tools for HWO science simulation},
  author       = {{Tumlinson}, Jason and {Riedel}, Adric },
  year = 2024,
  howpublished = {\url{https://github.com/spacetelescope/hwo-tools}},
  note         = {Accessed: 2025-08-28},
}
\end{document}